\documentclass[fleqn,usenatbib]{mnras}

\usepackage{newtxtext,newtxmath}

\usepackage[T1]{fontenc}
\usepackage{ae,aecompl}

\usepackage{graphicx}	
\usepackage{amsmath}	
\usepackage{amssymb}	

\title[Turbulence models in shock-cloud interactions]{A systematic comparison of two-equation RANS turbulence models applied to shock-cloud interactions}

\author[M. D. Goodson et al.]{
Matthew D. Goodson$^{1}$\thanks{E-mail: mgoodson@unc.edu}, Fabian Heitsch$^{1}$, Karl Eklund$^{2}$, and Virginia A. Williams$^{2}$
\\
$^{1}$Department of Physics \& Astronomy, University of North Carolina at Chapel Hill, Chapel Hill, NC 27599-3255, USA\\
$^{2}$Research Computing Center, University of North Carolina at Chapel Hill, Chapel Hill, NC 27599-3255, USA\\
}

\date{Accepted XXX. Received YYY; in original form ZZZ}

\pubyear{2017}

\begin{document}
\label{firstpage}
\pagerange{\pageref{firstpage}--\pageref{lastpage}}
\maketitle

\begin{abstract}
Turbulence models attempt to account for unresolved dynamics and diffusion in hydrodynamical simulations. We develop a common framework for two-equation Reynolds-Averaged Navier-Stokes (RANS) turbulence models, and we implement six models in the \textsc{Athena} code. We verify each implementation with the standard subsonic mixing layer, although the level of agreement depends on the definition of the mixing layer width. We then test the validity of each model into the supersonic regime, showing that compressibility corrections can improve agreement with experiment. For models with buoyancy effects, we also verify our implementation via the growth of the Rayleigh-Taylor instability in a stratified medium. The models are then applied to the ubiquitous astrophysical shock-cloud interaction in three dimensions. We focus on the mixing of shock and cloud material, comparing results from turbulence models to high-resolution simulations (up to 200 cells per cloud radius) and ensemble-averaged simulations. We find that the turbulence models lead to increased spreading and mixing of the cloud, although no two models predict the same result. Increased mixing is also observed in inviscid simulations at resolutions greater than 100 cells per radius, which suggests that the turbulent mixing begins to be resolved.
\end{abstract}

\begin{keywords}
hydrodynamics --- turbulence --- methods:numerical --- shock waves --- ISM:clouds
\end{keywords}



\section{Introduction}
The interstellar medium (ISM) is dominated by turbulent processes \citep{2004ARA&A..42..211E}. A common example is the interaction of a shock wave with a cloud of gas. Stellar winds and supernovae launch supersonic shock waves into the ISM that collide with nearby molecular gas clouds \citep{1977ApJ...218..148M}. The shock drives hydrodynamic instabilities at the cloud surface, such as the Rayleigh-Taylor (RT), Kelvin-Helmholtz (KH), and Richtmeyer-Meshkov (RM) instabilities, that disrupt and eventually destroy the cloud \citep{1992ApJ...390L..17S}. This interaction is well-studied in numerical simulations \citep{1992ApJ...390L..17S,1994ApJ...420..213K,1995ApJ...454..172X,2006ApJS..164..477N,2009MNRAS.394.1351P,2016MNRAS.457.4470P}.

In Eulerian hydrodynamics simulations, the growth of turbulence is controlled by numerical viscosity (resolution effects). Adequate resolution is therefore necessary to properly capture the dynamics. Previous work on the shock-cloud interaction has found that about 100 cells per radius are necessary for convergence of global quantities \citep{1994ApJ...420..213K,2006ApJS..164..477N,2009MNRAS.394.1351P}, although this requirement may be relaxed in 3D simulations \citep{2016MNRAS.457.4470P}. However, because the instabilities grow fastest on the smallest scales, the details of the small-scale mixing are dominated by resolution effects. \citet[][hereafter SSS08]{2008ApJ...680..336S} found that all quantities except the mixing fraction show convergence in shock-cloud simulations.

One possible means to mitigate resolution effects is a turbulence model, sometimes referred to as a subgrid-scale (SGS) model. Turbulence models attempt to mimic the effect of unresolved small-scale turbulence on the large-scale flow, often through the addition of ``turbulent'' stresses. Such models are common in engineering codes, and they are increasingly used in astrophysics \citep{2006A&A...450..265S,2008ApJ...686..927S,2009MNRAS.394.1351P,2011ApJ...733...88G,2011A&A...528A.106S,Schmidt2014b,2016MNRAS.457.4470P}. Turbulence models can be separated into two types: Reynolds-Averaged Navier-Stokes (RANS) and Large-Eddy Simulations (LES). The former relies on time-averaging of the decomposed fluid equations, while the latter uses spatial filtering of variables. Here, we only consider RANS models; for a review of LES methods, see \citet{Schmidt2014b}.

\subsection{Turbulence models in the shock-cloud interaction}\label{ss:prevshkcld}

Both RANS and LES turbulence models have been used to model the interaction of a shock with a cloud, in different environments and with different results. \citet[][hereafter P09]{2009MNRAS.394.1351P} examined the hydrodynamic shock-cloud interaction in two dimensions with the $k$-$\varepsilon$ model, a two-equation RANS model. The authors argued that the $k$-$\varepsilon$ turbulence model adequately captured the dynamics of the shock-cloud interaction and reduced the resolution requirements. Follow-up studies by \citet[][hereafter PP16]{2016MNRAS.457.4470P} revealed that the $k$-$\varepsilon$ model did not significantly alter the dynamics or improve the resolution convergence  in three dimensional simulations.

\citet[][hereafter GS11]{2011ApJ...733...88G} used a different two-equation RANS model, based on the $k$-$L$ formalism, to track metal enrichment in so-called ``minihalos''. An enriched supersonic galactic outflow impacts a diffuse cloud of primordial gas, subject to both gravity and radiative cooling. The authors modified the $k$-$L$ model of \citet[][hereafter DT06]{Dimonte2006}, which was calibrated for RT and RM instabilities, to include the KH instability and compressibility effects. Here the authors specifically investigated the turbulent mixing of metals. While there were notable differences in the enrichment of diffuse gas, the metal abundance in the dense gas was largely unaffected by the turbulence model.

\citet{2014MNRAS.440.3051S} applied a one-equation LES model to the simulations of \citet{2008MNRAS.388.1079I}, which studied a cosmological minor-merger, i.e., the infall of a low-mass subcluster into a larger cluster. This resembles the shock-cloud interaction but on larger scales. For this application, the authors used a linear eddy-viscosity relation with a dynamic procedure to calculate transport coefficients (``shear-improved'' SGS model). The authors found that, while the LES turbulence model did not significantly alter the energy of the interaction, it did affect the vorticity and subsequent evolution of the infalling gas.

It is difficult to interpret and compare the effects of the turbulence models in the simulations described above. First, each application explored different physical regimes and therefore included different physics (e.g. radiative cooling, gravity). Second, some turbulence models incorporated additional effects, such as buoyancy and compressibility, that other models implicitly neglect. Third, each turbulence model affects the dynamics differently. In the case of LES, the resolved dynamics are largely unaffected, as the model only considers turbulent effects near and below the filter width, which is typically close to the grid scale. However, RANS models average out dynamical fluctuations at all scales below some characteristic length scale, which varies throughout the simulation and could be much larger than the grid scale. Fourth, the ``true'' solution to the shock-cloud interaction is unknown. One can compare results obtained with a turbulence model to higher-resolution simulations, but without an explicit viscosity the degree of mixing remains constrained by the numerical viscosity.

Finally, it is unclear whether these turbulence models are valid in the astrophysical regimes being probed. All turbulence models rely on closure approximations with adjustable parameters often determined by comparison with empirical results. The laboratory experiments used for calibration are typically subsonic and incompressible in nature. While some models can be modified to produce correct results in transonic and moderately compressible regimes, it is unknown whether these modifications remain valid in the highly supersonic, highly compressible conditions characteristic of the ISM.

\subsection{Motivation and outline}\label{ss:motivation}

In an effort to better understand the effects and validity of turbulence models in astrophysical applications, we perform hydrodynamical simulations of the generic shock-cloud interaction with six two-equation RANS models. We first develop a common framework for two-equation turbulence models, and we implement this framework in the \textsc{Athena} hydrodynamics code \citep{2008ApJS..178..137S}. We verify the implementation of each turbulence model with the subsonic shear mixing layer test, ensuring that the width of the mixing layer grows linearly in accord with experimental results. We also highlight the dependence of the growth rate on the definition of the mixing layer width. We then test the validity of each model into the supersonic regime. Most models are known to perform poorly in transonic applications, but we explore three common ``compressibility corrections'' that improve results. Three of the models here considered include buoyancy effects, such as the RT instability. For these models, we further verify our implementation with a stratified medium test, in which we compare the temporal growth of the RT boundary layer to experimental results.

After determining that the turbulence models are implemented correctly, we test each turbulence model in a three-dimensional adiabatic shock-cloud interaction. We quantify not only the global dynamics but also the small-scale mixing. To examine the validity of the turbulence models, we perform a resolution convergence test of the inviscid shock-cloud interaction, up to 200 cells per radius in full 3D on a fixed grid. We also compare results to an ensemble-average of inviscid simulations initialized with grid-scale initial turbulence, scaled to roughly match the initial conditions of the turbulence models. Finally, we consider the effects of initial conditions and compressibility corrections in the turbulence models, finding that the former makes a significant difference in evolution whereas the latter does not.

We outline the six RANS turbulence models and their implementation in \textsc{Athena} in \S\ref{s:models}. We verify each implementation with a mixing layer test in \S\ref{s:mixinglayer}, and we further verify three of the models with the stratified medium test in \S\ref{s:rttest}. The turbulence models are then used in the shock-cloud simulation; the set-up and results of these simulations are presented and discussed in \S\ref{s:shkcloud}. Finally, we discuss the validity of turbulence models in astrophysical applications in \S\ref{s:discussion} before concluding in \S\ref{s:conclusions}.

\section{Turbulence models}\label{s:models}
We have modified the \textsc{Athena} hydrodynamics code \citep{2008ApJS..178..137S} version 4.2 to solve the system of equations:\footnote{For simplicity of notation, we do not differentiate Reynolds-averaged ($\overline{\rho},\overline{P}$) and Favre-averaged ($\tilde{\bf u}, \tilde{E}, \tilde{C}$) variables, where $\tilde{\phi} \equiv \overline{\rho \phi}/\overline{\rho}$.}
\begin{align}\label{eq:hydro}
  \frac{\upartial \rho}{\upartial t} + \nabla \cdot (\rho {\bf u}) &= 0 \\
  \frac{\upartial (\rho {\bf u})}{\upartial t} + \nabla \cdot (\rho {\bf u} {\bf u} + P{\bf I}) &=  \nabla \cdot  {\boldmath \tau'} \\
  \frac{\upartial E}{\upartial t} + \nabla \cdot [(E+P) {\bf u}] &= \nabla \cdot ({\bf u} {\bf \tau'} - {\bf q'}) + \Psi_E \\
  \frac{\upartial (\rho C)}{\upartial t} + \nabla \cdot (\rho C {\bf u}) &= \nabla \cdot {\bf d'}\\
  \frac{\upartial (\rho k)}{\upartial t}  + \nabla \cdot (\rho k {\bf u}) &= \nabla \cdot (\frac{\mu_T}{\sigma_k} \nabla k) + \Psi_k \label{eq:dkteq} \\
  \frac{\upartial (\rho \xi)}{\upartial t}  + \nabla \cdot (\rho \xi {\bf u}) &= \nabla \cdot (\frac{\mu_T}{\sigma_\xi} \nabla \xi) + \Psi_\xi \label{eq:dxieq}
\end{align} with the density $\rho$, the fluid velocity vector ${\bf u}$, the pressure $P$, the unit dyad ${\bf I}$, the total resolved energy density $E$ \footnote{We do not include the turbulent kinetic energy $\rho k$ in the definition of total energy; therefore we are simulating the total {\em resolved} energy. See section 2.4.5 of \citet{Garnier2009} for a complete discussion of compressible energy equation systems.}:
\begin{equation}\label{eq:eos}
E = \frac{P}{\gamma -1} + \frac{1}{2} \rho |{\bf u}|^2,
\end{equation}a passive colour field $C$, the specific turbulent kinetic energy $k$, an auxiliary turbulence variable $\xi$, the turbulent stress tensor ${\boldmath \tau'}$, the turbulent heat flux ${\bf q'}$, the turbulent diffusive flux ${\bf d'}$, turbulent viscosity $\mu_T$, turbulent diffusion coefficients $\sigma$, and source terms due to turbulent effects $\Psi$.

Two-equation models are so named because they add two ``turbulent'' variables -- the specific turbulent kinetic energy $k$ and an auxiliary variable $\xi$ that varies from model to model -- with corresponding transport equations (Eqs. \ref{eq:dkteq}-\ref{eq:dxieq}). Models are typically denoted by the chosen auxiliary turbulence variable; e.g., $\xi \to \varepsilon$ yields the $k$-$\varepsilon$ model. Here, we examine the standard $k$-$\varepsilon$ model of \citet[][hereafter LS74]{Launder1974}, as well as the extended model of \citet[][hereafter MS13]{Moran-Lopez2013a}; the $k$-$L$ models of \citet[][hereafter C06]{Chiravalle2006} and GS11; and the $k$-$\omega$ models of \citet[][hereafter W88]{Wilcox1988} and \citet[][hereafter W06]{Wilcox2006}. For the $k$-$\varepsilon$ and $k$-$\omega$ models, we also test the effect of three standard compressibility corrections, presented in \citet[][hereafter S89]{Sarkar1989}, \citet[][hereafter Z90]{Zeman1990}, and \citet[][hereafter W92]{Wilcox1992}.

The turbulent stress tensor ${\boldmath \tau'}$ is defined as
\begin{equation}
  \tau'_{ij} = 2 \mu_T (S_{ij} - \frac{1}{3}\delta_{ij}S_{kk}) - \frac{2}{3} \delta_{ij} \rho k
\end{equation}with resolved stress rate tensor ${\bf S}$ given by
\begin{equation}\label{eq:sij}
S_{ij} = \frac{1}{2}(\frac{\upartial u_i}{\upartial x_j} + \frac{\upartial u_j}{\upartial x_i}).
\end{equation}The specific turbulent kinetic energy $k$ is defined as $k \equiv (1/2) \tau'_{kk}$ and requires an additional transport equation. The generic transport equation (Eq. \ref{eq:dkteq}) is applicable to (almost) all models investigated, with source term
\begin{equation}\label{eq:psik}
\Psi_k = P_T - C_D \rho \varepsilon + C_B \rho \sqrt{k} A_i g_i
\end{equation}with the production term $P_T = \tau'_{ij} \upartial u_i/\upartial x_j$, specific dissipation $\varepsilon$, dissipation coefficient $C_D$, buoyancy coefficient $C_B$, and Atwood number in the $i$th direction $A_i$ with acceleration $g_i = {-(1/\rho)\upartial P / \upartial x_i}$. The source term on the energy equation is $\Psi_{E} = -\Psi_{k}$. Table \ref{tab:constants} presents a summary of all model constants and values.

In adiabatic simulations, the turbulent heat flux vector ${\bf q'}$ is defined as
\begin{equation}
q'_j = -\kappa_T \frac{\upartial T}{\upartial x_j} = \frac{\gamma}{\gamma-1} \frac{\mu_T}{{Pr}_T} \frac{\upartial T}{\upartial x_j}
\end{equation}with turbulent thermal conductivity $\kappa_T = c_p \mu_T / {Pr}_T$, specific heat capacity $c_p = \gamma/(\gamma-1)$, and turbulent Prandtl number ${Pr}_T$.

Passively advected scalar fields are diffused using a gradient-diffusion approximation, where the turbulent diffusive flux vector ${\bf d'}$ is given by
\begin{equation}
d'_j = \frac{\mu_T}{\sigma_C}  \frac{\upartial C}{\upartial x_j},
\end{equation}with Schmidt number $\sigma_C$ generally of order unity.

\begin{table*}
  \centering
  \caption{Summary of model constants. Some values may appear at variance with the reference; this is due only to our generic formalism, which redefines and combines certain constants for consistency across all models. Values presented for the W06 model neglect limiting functions and should therefore be considered approximations.}
  \label{tab:constants}
  \begin{tabular}{llcccccc}
    \hline
Constant         & Description               & LS74 & MS13 &  C06 & GS11 &  W88 &  W06 \\
    \hline
$C_\mu$          & Turbulent viscosity       & 0.09 & 0.09 & 0.30 & 1.00 & 1.00 & 1.00 \\
$C_D$            & Dissipation of turbulence & 1.00 & 1.00 & 8.91 & 3.54 & 0.09 & 0.09 \\
$C_B$            & Buoyancy effects          & 0.00 & 0.10 & 1.70 & 1.19 & 0.00 & 0.00 \\
$Pr_T$           & Turbulent Prandtl number  & 0.90 & 0.90 & 1.00 & 1.00 & 0.90 & 0.89 \\
$\sigma_k$       & Turbulent energy diffusion& 1.00 & 0.50 & 1.00 & 1.00 & 2.00 & 1.67 \\
$\sigma_\xi$     & Turbulent diffusion       & 1.30 & 0.50 & 1.00 & 0.50 & 2.00 & 2.00 \\
$\sigma_C$       & Turbulent Schmidt number  & 1.00 & 0.50 & 1.00 & 1.00 & 1.00 & 1.00 \\
$C_1$            & Turbulence generation     & 1.44 & 1.44 & 1.00 & 0.33 & 0.56 & 0.52 \\
$C_2$            & Additional effects        & 1.92 & 1.92 & 1.00 & 1.00 & 0.08 & 0.07 \\
$C_3$            & Buoyancy effects          & 0.00 & 0.09 & 0.00 & 0.00 & 0.00 & 0.00 \\
    \hline
  \end{tabular}
\end{table*}

\subsection{\texorpdfstring{$k$-$\varepsilon$ models}{k-epsilon models}}\label{ss:keps}
In the $k$-$\varepsilon$ formalism, the auxiliary turbulence variable $\xi$ is defined to be the specific turbulent energy dissipation $\varepsilon \propto k^{3/2} L^{-1}$, where $L$ is a defined turbulent length scale. The exact scaling depends on the implementation; we here use $\varepsilon = C_\mu^{3/4} k^{3/2} L^{-1}$, where $C_\mu$ is a model constant related to the viscosity.

\subsubsection{LS74}\label{sss:LS74}
LS74 outlined the standard version of the $k$-$\varepsilon$ model, and it is perhaps the most widely used RANS turbulence model. The model uses the eddy-viscosity $\mu_T$ defined as
\begin{equation}
  \mu_T = C_\mu \rho \frac{k^2}{\varepsilon}
\end{equation} with $C_\mu = 0.09$. The transport equation for $\varepsilon$ (Eq. \ref{eq:dxieq}) has the source term
\begin{equation}
  \Psi_\varepsilon = C_1 \frac{\varepsilon}{k} P_T - C_2 \rho \frac{\varepsilon^2}{k}.
\end{equation}The model constants are summarized in Table \ref{tab:constants}. Because $C_B = 0$, the model neglects buoyant effects, such as the RT instability.

\subsubsection{MS13}\label{sss:MS13}
To include the RT and RM instability effects in the $k$-$\varepsilon$ model, MS13 added a buoyancy term, with the Atwood number in Eq. \ref{eq:psik} defined as
\begin{equation}\label{eq:msa}
A_i = \frac{k^{3/2}}{\rho \varepsilon} (\frac{\upartial \rho}{\upartial x_i} - \frac{\rho}{P}\frac{\upartial P}{\upartial x_i}).
\end{equation} The source term for the dissipation equation $\Psi_\varepsilon$ is also extended as
\begin{equation}\label{eq:psie}
  \Psi_\varepsilon = C_1 \frac{\varepsilon}{k} P_T - C_2 \rho \frac{\varepsilon^2}{k} + C_3 \rho \frac{\varepsilon}{\sqrt{k}} A_i g_i.
\end{equation}The model constants are summarized in Table \ref{tab:constants}; we note that the MS13 values are largely the same as LS74 but with modified transport coefficients and $C_B \neq 0$.

\subsection{\texorpdfstring{$k$-$L$ models}{k-L models}}\label{ss:kl}
The $k$-$L$ model is a two-equation RANS model developed by DT06 to study RT and RM instabilities. Shear (KH instability) was added by C06 and extended to include compressibility effects by GS11. The auxiliary variable $\xi$ is defined to be the eddy length scale $L$. The model uses the eddy-viscosity
\begin{equation}\label{eq:klmut}
  \mu_T = C_\mu \rho L \sqrt{2 k}.
\end{equation}The transport equation for $L$ (Eq. \ref{eq:dxieq}) has the source term
\begin{equation}\label{eq:psil}
  \Psi_L = C_1 \rho L (\nabla \cdot {\bf u}) + C_2 \rho \sqrt{2 k}.
 \end{equation}Again, we here set the specific dissipation in Eq. \ref{eq:psik} to be $\varepsilon = C_\mu^{3/4} k^{3/2} L^{-1}$.

\subsubsection{C06}\label{sss:C06}
C06 added shear to the $k$-$L$ model of DT06 by employing the full stress tensor rather than just the turbulent pressure term. This necessitated re-calibrating the model coefficients of DT06. We note that C06 used a slightly different RT growth rate parameter ($\alpha = 0.05$ instead of $\alpha = 0.0625$ in DT06) when calibrating the model. Buoyancy effects are included via the Atwood number defined as
\begin{equation}
A_i = \frac{\rho_+ - \rho_-}{\rho_+ + \rho_-} + \frac{L}{\rho}\frac{\upartial \rho}{\upartial x_i},
\end{equation}where $\rho_+$ and $\rho_-$ are the reconstructed density values at the right and left cell faces, respectively. The model constants are summarized in Table \ref{tab:constants}; we note that the constant values appear to differ from those given in C06, but that this is solely due to our generic two-equation framework which combines and re-defines certain constants.

\subsubsection{GS11}\label{sss:GS11}
Similar to C06, the model of GS11 is based on the $k$-$L$ model of DT06, but with the complete turbulent stress tensor to include KH effects. The model uses a slightly different definition of the Atwood number from C06, with
\begin{equation}
A_i = \frac{\rho_+ - \rho_-}{\rho_+ + \rho_-} + \frac{2 L}{\rho + L\,|\upartial \rho/\upartial x_i|}\frac{\upartial \rho}{\upartial x_i},
\end{equation}where again $\rho_+$ and $\rho_-$ are the reconstructed density values at the right and left cell faces, respectively.

GS11 also introduces a variable ($\tau_{\rm KH}$) to account for compressibility effects by modifying the turbulent stress tensor,
\begin{equation}
  \tau'_{ij} = 2 \mu_T \tau_{\rm KH} (S_{ij} - \frac{1}{3}\delta_{ij}S_{kk}) - \frac{2}{3} \delta_{ij} \rho k.
\end{equation}$\tau_{\rm KH}$ is calibrated with compressible shear layer simulations and estimated using a ``local'' Mach number $M_l \equiv |\nabla \times {\bf u}|\, L / c_s$, where $c_s$ is the local sound speed. However, the piecewise fit for $\tau_{\rm KH}$ given by Eq. 19 in GS11 is discontinuous, which can lead to numerical issues. We therefore fit their formulation with a smooth function,
\begin{equation}\label{eq:tauKH}
\tau_{\rm KH}(M_l) = 0.000575 + \frac{0.19425}{1.0 + 0.000337 {\rm exp}(17.791~M_l)}.
\end{equation}The model constants are summarized in Table \ref{tab:constants}; we note that the C06 and GS11 model constants differ despite significant similarity in model formulation and calibration.

\subsection{\texorpdfstring{$k$-$\omega$ models}{k-omega models}}\label{ss:komega}
The $k$-$\omega$ model was first developed by W88 and updated in \citet{Wilcox1998} and W06. The auxiliary variable $\xi$ is defined to be the specific dissipation rate (or eddy frequency) $\omega = k^{1/2} L^{-1}$, which has units of inverse time. Then the specific dissipation is $\varepsilon = C_\mu k \omega$. To our knowledge, this is the first use of a $k$-$\omega$ model in an astrophysical application.

\subsubsection{W88}\label{sss:W88}
The first version of the $k$-$\omega$ model is outlined in W88. The model uses the eddy-viscosity
\begin{equation}
  \mu_T = C_\mu \frac{\rho k}{\omega}.
\end{equation}The transport equation for $\omega$ (Eq. \ref{eq:dxieq}) uses the source term
\begin{equation}\label{eq:psiw}
  \Psi_\omega = C_1 \frac{\omega}{k} P_T - C_2 \rho \omega^2.
\end{equation}The model constants are summarized in Table \ref{tab:constants}.

\subsubsection{W06}\label{sss:W06}
The most recent version of the $k$-$\omega$ model is presented in W06 and \citet{Wilcox2008}. While the model is similar to W88, there are important (and elaborate) differences, such as cross-diffusion terms and stress limiters. While the additional terms improve the accuracy and reduce the dependence on initial conditions, the model is sufficiently complex to prohibit a generic description. Our implementation in \textsc{Athena} includes the additional terms, and we refer the reader to W06 and \citet{Wilcox2008} for a full description of the model. For completeness we note approximate constant values in Table \ref{tab:constants}.

\subsection{Compressibility Corrections}\label{ss:compcorr}
A common way to account for compressibility effects is to modify the turbulence dissipation rate $\varepsilon$. In theory, $\varepsilon$ is decomposed into solenoidal and dilatational components, with the latter only manifesting in compressible turbulence. In practice, only a slight modification is needed to the $k$ and $\omega$ equations. In Eq. \ref{eq:psik}, the second term on the right hand side is modified as $C_D \rho \varepsilon \to C_D \rho \varepsilon [1 + F(M_t)]$, where $F(M_t)$ is a function of the local turbulent Mach number $M_t \equiv \sqrt{2 k}/a_s$, with $a_s$ the local sound speed. No further changes are needed in the $k$-$\epsilon$ formalism. In the $k$-$\omega$ formalism, Eq. \ref{eq:psiw} is also modified with $C_2 \rho \omega^2 \to [C_2 - C_D F(M_t)] \rho \omega^2$. We consider three forms for $F(M_t)$ proposed in the literature. The simplest model is that of S89 which uses
\begin{equation}
F(M_t) = M_t^2.
\end{equation}The most complex model is that of Z90 with
\begin{equation}
F(M_t) = 0.75 \{1.0 - {\rm exp}[-1.39( \gamma + 1.0)(M_t - M_{t0})^2]\} \mathcal{H}(M_t - M_{t0}),
\end{equation} with $\mathcal{H}$ the Heaviside step function and $M_{t0} \equiv 0.10 \sqrt{2/(\gamma + 1)}$. Finally, the model of W92 suggests
\begin{equation}
F(M_t) = 1.5 (M_t^2 - 0.0625) \mathcal{H}(M_t - 0.25).
\end{equation} It is worth noting that these are purely phenomenological models; resolved DNS simulations by \citep{Vreman1996} have demonstrated that the dissipation is not actually reduced in compressible turbulence. Despite this realization, compressibility corrections that modify the dissipation are still commonly used because they yield accurate results in many applications. As noted in \S\ref{sss:GS11}, GS11 uses a different type of compressibility correction which modifies the turbulent stress tensor. No satisfactory correction is available for C06.

\subsection{Turbulence model initial conditions}\label{ss:turbinit}
In simulations with a turbulence model, we must specify initial conditions for the turbulent kinetic energy $k$ and the additional turbulent variable ($\xi \to \varepsilon$, L, or $\omega$). We desire identical initial conditions for all models; we therefore set the turbulent length scale $L$ in all models and convert using scaling relations. Based on dimensional arguments, $\varepsilon \propto k^{3/2} / L$ and $\omega \propto k^{1/2} / L$. The literature values for the constant of proportionality vary; we obtained the best agreement across models using $\varepsilon_0 = C_\mu^{3/4} k_0^{3/2} L_0^{-1}$ and $\omega_0 = C_\mu^{-1/4} k_0^{1/2} L_0^{-1}$.

\subsection{Implementation in \textsc{Athena}}\label{ss:implementation}
The turbulence update is first order in time and implemented via operator splitting. The fluxes are calculated at cell walls using a simple average to reconstruct quantities from cell-centred values. Spatial derivatives are computed using second order central differences. Source terms are evaluated after application of the viscous fluxes and are applied with an adaptive Runge-Kutta-Fehlberg integrator (RKF45). Stability of the explicit diffusion method is preserved by limiting the overall hydrodynamic time step based on the condition $\Delta t \le  (\Delta^2 \rho)/ (6\mu_T)$, where $\Delta$ is the minimum cell size. The dependence on $\Delta^2$ limits the feasibility of our implementation to low resolution simulations.

\section{Mixing layer test}\label{s:mixinglayer}

\begin{table*}
  \centering
  \caption{Mixing layer growth rates for $M_{\rm c}$ = 0.10.}
  \label{tab:shearresults}
  \begin{tabular}{lccccc}
    \hline
    Model & $C_{b10}$ & $C_{b1}$ & $C_{s10}$ & $C_{\omega}$ & $C_{\theta}$ \\
    \hline
empirical  & 0.082-0.100 $^{a}$
           & 0.170-0.181 $^{b}$
           & 0.058-0.084 $^{a}$
           & 0.081-0.091 $^{a,c}$
           & 0.016-0.018 $^{d}$   \\
LS74       & 0.070
           & 0.092
           & 0.056
           & 0.083
           & 0.015 \\
MS13       & 0.067
           & 0.091
           & 0.053
           & 0.077
           & 0.014 \\
C06        & 0.181
           & 0.206
           & 0.143
           & 0.066
           & 0.038 \\
GS11       & 0.123
           & 0.189
           & 0.100
           & 0.134
           & 0.026 \\
W88        & 0.052
           & 0.062
           & 0.041
           & 0.040
           & 0.011 \\
W06        & 0.061
           & 0.074
           & 0.047
           & 0.069
           & 0.013 \\
    \hline
  \end{tabular}\\
$^{a}$\citet{Barone2006};
$^{b}$\citet{Papamoschou1988}, with $\delta_{viz} \approx \delta_{b1}$;
$^{c}$\citet{Brown1974};
$^{d}$\citet{Pantano2002}.
\end{table*}

To verify the implementation of each turbulence model in \textsc{Athena}, we perform a one-dimensional temporal mixing layer test. Our set-up is nearly identical to that described in section 2.2.2 of GS11, which was adapted from section 3 of C06. We initialize a discontinuity in the perpendicular ($y$) velocity at the origin. The difference in velocity between the left and right states sets the convective Mach number, defined as \citep{Papamoschou1988}
\begin{equation}
M_{\rm c} \equiv \frac{|v_l - v_r|}{c_l + c_r},
\end{equation}with $v$ the $y$-velocity and $c$ the sound speed, with subscripts $l$ and $r$ for the left and right regions respectively. Unlike GS11, we shift the frame of reference to move at the convective velocity; then $v_l = -v_r$. We also smooth the initial velocity discontinuity with a hyperbolic tangent function, as was done in \citet{2008ApJ...678..234P}. The parallel ($x$) velocity is zero. We use an ideal equation of state with $\gamma = 1.4$. The density and pressure are constant at $\rho_0 = 1.0\,{\rm g}\,{\rm cm}^{-3}$ and $P_0=1.72\times10^{10}\,{\rm erg}\,{\rm cm}^{-3}$, corresponding to a uniform sound speed $c_l = c_r = 1.55\times10^5\,{\rm cm}\,{\rm s}^{-1}$. The simulation domain is a one-dimensional region with extent -5.0\,cm $< x <$ 5.0\,cm with a resolution of 4096 cells. Similar to GS11, we initialize a small shear layer of width $\delta_0 = 0.1\,{\rm cm}$ centred at the interface with turbulent energy $k = 0.02 (\Delta v)^2$ and $L = 0.2 \delta_0$, where $\Delta v = |v_l - v_r|$. This initial layer is also smoothed to the background values of $k_0 = 10^{-4} (\Delta v)^2$ and $L_0 = 10^{-2} \delta_0$.

We run each simulation for 200$\,\mu$s. The velocity discontinuity generates a shear layer, and the width of the shear layer $\delta$ grows linearly in time as
\begin{equation}
  \delta(t) = C_\delta~\Delta v~t,
\end{equation}where $C_\delta$ is a constant. The exact value for $C_\delta$ depends on how the shear layer thickness $\delta$ is defined. In lab experiments, the visual thickness $\delta_{\rm viz}$ \citep{Brown1974} or pressure thickness $\delta_p$ \citep{Papamoschou1988} are used. In numerical experiments, the velocity thickness $\delta_b$, energy thickness $\delta_s$, and vorticity thickness $\delta_\omega$ are often used \citep{Barone2006}; less common is the momentum thickness, $\delta_\theta$ \citep{Vreman1996}. C06 and GS11 used a 1 per cent threshold on the velocity thickness (which we will denote as $\delta_{b1}$), considering regions where $0.01 < (v-v_l)/(\Delta v) < 0.99$; engineering literature tends to use a 10 per cent threshold ($\delta_{b10}$), defined similarly to $\delta_{b1}$. W88 used a 10 per cent energy thickness ($\delta_{s10}$), defined where $0.1 < (v-v_l)^2/(\Delta v)^2 < 0.9$. We will compare results using these three definitions, as well as the momentum thickness $\delta_\theta = 1/[\rho_0 (\Delta v)^2] \int \rho (v_l - v)(v - v_r)~dx$ and the vorticity thickness $\delta_\omega = |v_l - v_r|/(\upartial v/\upartial y)_{max}$.

A further complication is that lab experiments of the plane mixing layer measure a spatial spreading rate, $\delta'(x) \equiv d \delta / d x$. In our experiment, we move in a frame of reference at the convective velocity $v_{\rm c} = (1/2)(v_l+v_r)$ (assuming $c_l = c_r$) and therefore measure a temporal spreading rate, \citep[e.g.,][]{Vreman1996,Pantano2002}
\begin{equation}
\delta'(t) = \frac{d \delta}{d t} = \frac{d x}{d t} \frac{ d \delta}{d x} = v_c \delta'(x).
\end{equation}Values for $C_\delta$ estimated from plane mixing layer experiments \citep{Brown1974,Papamoschou1988} and high-resolution numerical simulations \citep{Pantano2002,Barone2006} are reported in Table \ref{tab:shearresults}, where the subscript on $C$ indicates the corresponding shear layer thickness definition.

\subsection{Mixing layer results}\label{ss:mlresults}

\begin{figure}
  \includegraphics[width=\linewidth]{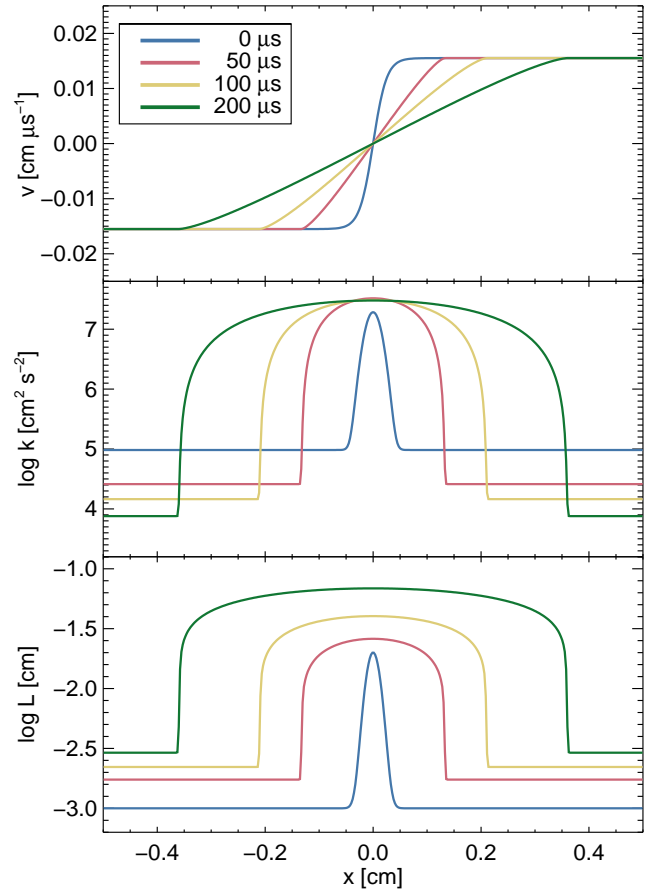}
  \caption{Time evolution of the one-dimensional subsonic ($M_{\rm c} = 0.10$) shear flow test with the LS74 $k$-$\varepsilon$ turbulence model. From the top, profiles of the $y$-velocity $v$, specific turbulent kinetic energy $k$, and turbulent length scale $L = C_\mu^{3/4} k^{3/2} \varepsilon^{-1}$. Profiles are shown at times $t = $ 0, 50, 100, and 200 $\mu$s, indicated by colour.}
  \label{f:chiravalle_profiles}
\end{figure}

\begin{figure}
  \includegraphics[width=\linewidth]{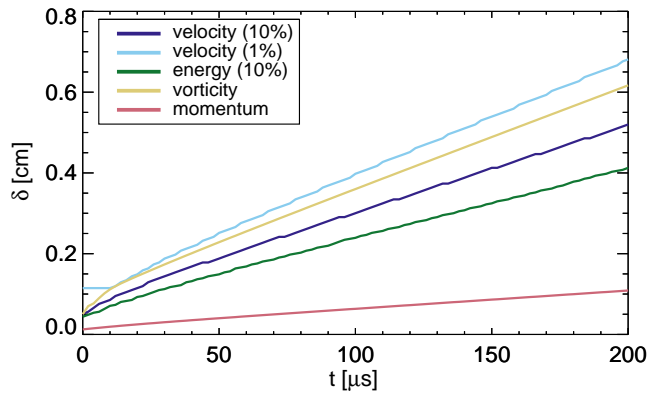}
  \caption{Growth of the shear layer width $\delta(t)$ in the subsonic ($M_{\rm c} = 0.10$) mixing layer with the LS74 $k$-$\varepsilon$ turbulence model. The shear layer definition is indicated by colour. All definitions produce linear growth but at different rates.}
  \label{f:chiravalle_growth}
\end{figure}

Figure \ref{f:chiravalle_profiles} shows the time evolution of a subsonic ($M_{\rm c} = 0.1$) mixing layer with the LS74 $k$-$\varepsilon$ model. The profiles of the the $y$-velocity $v$, turbulent kinetic energy $k$, and turbulent length $L$ all spread in time; as noted, the exact spreading rate depends on how the layer thickness is defined. Figure \ref{f:chiravalle_growth} shows the growth of the shear layer thickness $\delta(t)$ for different layer definitions. All definitions show linear growth in time. The 1 per cent velocity thickness grows at the greatest rate, while the momentum thickness increases at the lowest rate. We use a $\chi^2$ minimization linear fit to estimate $C_\delta$; the results are presented in Table \ref{tab:shearresults}.

Table \ref{tab:shearresults} also shows the growth rates at $M_{\rm c} = 0.10$ for all RANS models tested. We find that the various turbulence models lead to differing growth rates on the same test problem. Although most models do not reproduce the measured growth rate for all thickness definitions, all models do produce linear growth in time and roughly agree with the measured value for at least one definition, leading us to conclude that our models are implemented correctly in \textsc{Athena}. Variations in numerical method between codes could lead to discrepancies with previous work; further, there is significant uncertainty on the measured values. Interestingly, there is no clear relation between the different measures and models; for example, $C_{b10}$ is much greater with the GS11 model compared to the LS74 model, but $C_\omega$ is slightly less. This suggests no single measure should be preferred.

Finally, we note that C06 and GS11 calibrated their turbulence models using a 1 per cent velocity definition for the mixing layer. While their models show good agreement with this definition, we find that these models largely do not predict spreading rates in agreement with measured values when using other definitions. This suggests that a 1 per cent criterion may not be the best definition for comparison.

\subsection{Compressible mixing layer}\label{ss:mixcompcorr}

\begin{figure}
  \includegraphics[width=\linewidth]{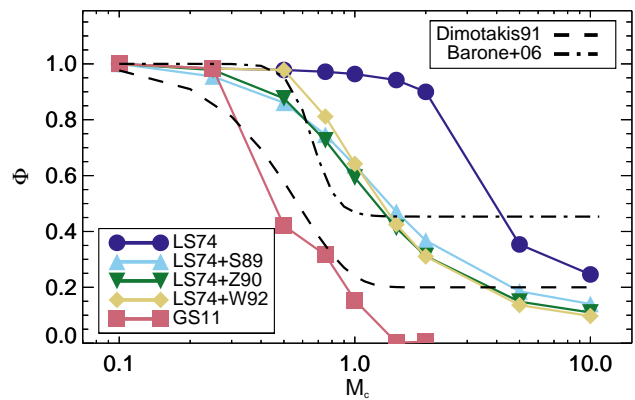}
  \caption{The compressibility factor $\Phi \equiv \delta'/\delta'_i$ as a function of convective Mach number $M_{\rm c}$. Results are shown for the standard LS74 model with no compressibility correction (purple dots) and with the compressibility corrections of S89 (blue upward triangles), Z90 (green downward triangles), and W92 (gold diamonds); as well as for the GS11 model (red squares), which includes a stress modification ($\tau_{\rm KH}$). The empirical curves of \citet[][dashed]{Dimotakis1991} and \citet[][dot-dashed]{Barone2006} are also shown for comparison.}
  \label{f:chiravalle_comp}
\end{figure}

The spreading rate of a compressible mixing layer is found to decrease with increasing convective Mach number \citep{Birch1972,Brown1974,Papamoschou1988}. The difference is expressed as the compressibility factor $\Phi \equiv \delta'/\delta'_i$, where $\delta'_i$ is the incompressible growth rate. Experiments have yielded different relations between $M_c$ and $\Phi$, such as the popular ``Langley'' curve \citep{Birch1972}, the results of \citet{Papamoschou1988}, and the fit of \citet{Dimotakis1991}.

We perform simulations with increasing convective Mach number up to $M_c = 10$. We use the growth rate determined at $M_c = 0.1$ with thickness $\delta_{b10}$ as our incompressible growth rate $\delta'_i$. Results obtained with the LS74 model are presented as solid circles in Figure \ref{f:chiravalle_comp}, with two experimental curves shown for comparison. Although the spreading rate does decrease with increasing Mach number, it does not follow the experimental trend. This is consistent with previous work which shows that standard two-equation RANS turbulence models do not reproduce the observed reduction in spreading rate without modifications.

As described in \S\ref{ss:compcorr}, three authors (S89, Z90, and W92) have proposed ``compressibility corrections'' to better capture the decrease. These corrections work by increasing the dissipation rate due to pressure-dilatation effects. Although direct numerical simulation results have shown that this is not actually the case \citep{Vreman1996}, these \textit{ad hoc} compressibility corrections are still widely used because they produce more accurate results (at least in the transonic regime). Figure \ref{f:chiravalle_comp} also shows results obtained when the three compressibility corrections are applied to the LS74 model. All three corrections do decrease the spreading rate to roughly the experimental values, at least up to $M_c = 5$; above this, the growth rate is slightly below the experimental estimate. The difference between the corrections of S89, Z90, and W92 is negligible. Similar results are obtained when applied to the MS13, W88, and W06 models.

There is no straightforward way to apply these corrections to the model of C06; however, GS11 does include a compressibility correction through the variable $\tau_{\rm KH}$ (see Section \ref{sss:GS11}). Results obtained with the model of GS11 are also shown on Figure \ref{f:chiravalle_comp}. The asymptotic nature of the $\tau_{\rm KH}$ function (Eq. \ref{eq:tauKH}) reproduces the observed behavior of compressible layers up to $M_c \approx 1$; however, above this point the GS11 formulation leads to growth rates that are too small. Indeed, data points are not available for $M_c > 2.5$ for GS11 because the model did not evolve.

\section{Stratified medium test}\label{s:rttest}

\begin{figure}
  \includegraphics[width=\linewidth]{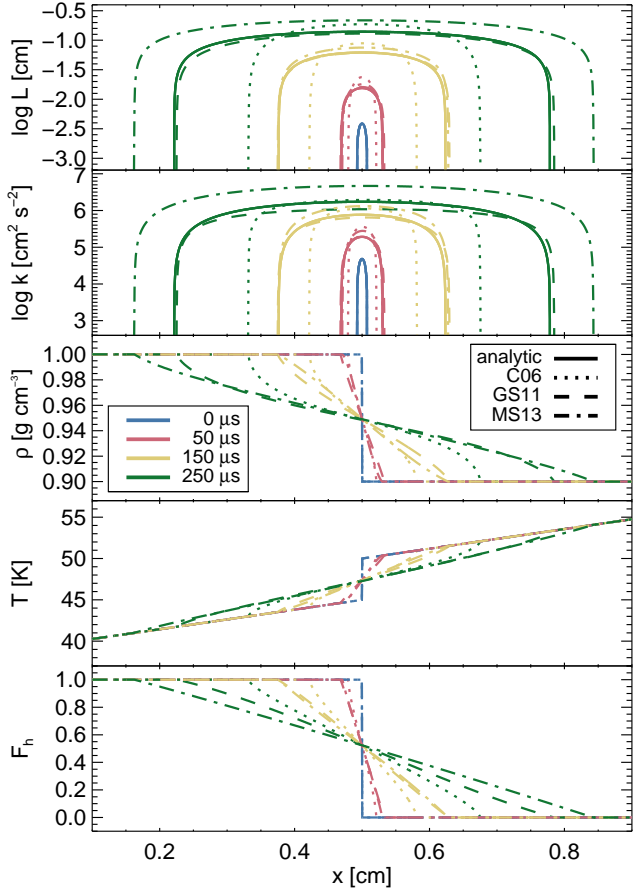}
  \caption{Time evolution of the stratified medium test with the buoyant turbulence models (dotted: C06; dashed: GS11; dot-dashed: MS13). From the top, profiles of the turbulent length scale $L$, specific turbulent kinetic energy $k$, density $\rho$, temperature $T$, and heavy-fluid mass fraction $F_{\rm h}$. Profiles are shown at times $t = $ 50, 100, 200, and 300 $\mu$s, indicated by colour. Analytic solutions are shown for $L$ and $k$ with solid lines. While GS11 matches well, C06 grows too slowly and MS13 too quickly.}
  \label{f:dimonte_profiles}
\end{figure}

\begin{figure}
  \includegraphics[width=\linewidth]{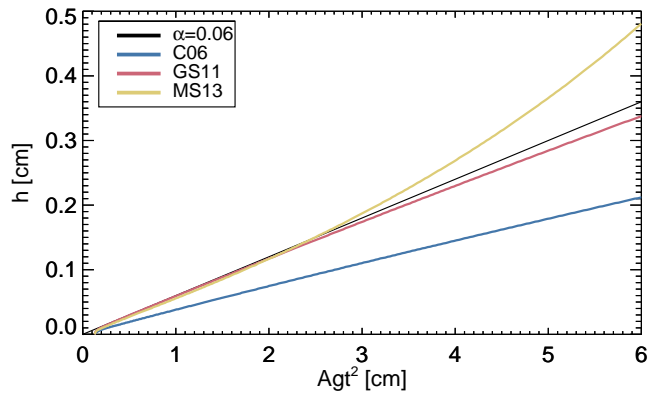}
  \caption{Growth of the bubble height $h$ as a function of $A g t^2$ for the buoyant turbulence models (C06, GS11, and MS13). The growth should be linear with slope equal to the DT06 experimental bubble constant $\alpha = 0.06$, shown in black. GS11 matches well with linear growth at $\alpha \approx 0.05$. C06 is also linear but with a lower value of $\alpha \approx 0.038$. MS13 matches well initially with $\alpha \approx 0.06$, but eventually the evolution becomes non-linear and diverges.}
  \label{f:dimonte_growth}
\end{figure}

Three of the models here considered include buoyant effects to capture the RT instability, namely MS13, C06, and GS11. To further verify the implementation of these models, we perform a two-dimensional stratified medium test. Our set-up is nearly identical to that described in section 2.2.1 of GS11, which was itself adapted from section 5 of DT06. We accelerate a heavy fluid of density $\rho_1 = 1.0$~g~cm$^{-3}$ into a lighter fluid of density $\rho_2 = 0.9$~g~cm$^{-3}$ from an initially hydrostatic state. The acceleration acts in the $-y$ direction at $g = 9.8\times10^8$~cm~s$^{-2}$. The grid is $0.02\times1.0$~cm with a resolution of $16\times800$ cells, and the interface is at the midpoint of the $y$ axis. The temperature is discontinuous at the interface, with $T_1 = 45$~K and $T_2 = 50$~K, and follows a profile to maintain hydrostatic equilibrium. Note that we do not perturb the interface; as the interface is grid-aligned, the RT instability will not develop in an inviscid code. However, a buoyant turbulence model will recognize the impulsive density and pressure gradients and generate turbulence, leading to the development of a mixing layer between the two fluids. Bubbles of light fluid will penetrate the heavy fluid with height $h(t) = \alpha_b A g t^2$, where $A = (\rho_1 - \rho_2)/(\rho_1 + \rho_2)$ is the Atwood number and $\alpha_b \approx 0.06$ is a constant empirically determined from experiments \citep{Dimonte2004}. Numerical simulations of the RT instability tend to underestimate the growth by a factor of $\sim 2$ \citep{Dimonte2004,2007PhFl...19i4104S}, underscoring the need for a turbulence model.

Figure \ref{f:dimonte_profiles} shows the evolution of the boundary layer for the turbulence models of C06, GS11, and MS13. The other turbulence models (LS74, W88, and W06) lack buoyant source terms; hence they cannot capture the RT instability and show no evolution in this test case. We compare the growth of turbulent kinetic energy $k(y,t)$ and turbulent length scale $L(y,t)$ with the analytic solutions given in DT06. The model of GS11 shows good agreement with the analytic predictions; however, the models of C06 and MS13 do not accurately follow the evolution. We note that C06 used a slightly lower value of the bubble penetration constant $\alpha_b$ compared to DT06 when calibrating the model; however this is insufficient to fully explain the discrepancy. Figure \ref{f:dimonte_profiles} also shows the evolution of the density $\rho$, the temperature $T$, and the heavy fluid mass fraction $F_{\rm h}$, determined using a passive colour field $C$ that is initialized to unity in the heavy fluid and to zero in the light fluid.

We can also determine the growth rate of the bubble height $h(t)$, estimated as the point where the mass fraction of heavy material $F_{\rm h} = 0.985$ \citep{2007PhFl...19i4104S}. Figure \ref{f:dimonte_growth} shows the growth of the bubble height $h(t)$ plotted against $Agt^2$; hence the lines should be linear with a slope of $\alpha \approx 0.06$. We see that, after an initial transient phase, the GS11 model does show a linear trend with $\alpha \approx 0.050$ -- slightly lower than expected but still in good agreement. The model of C06 also shows a linear trend, but the layer grows too slowly with $\alpha \approx 0.038$. The MS13 model is initially in good agreement with $\alpha \approx 0.060$ but eventually diverges and grows non-linearly. It is unclear what in the MS13 model causes this runaway growth, but the test result suggests that MS13 may not properly account for sustained buoyancy and will therefore yield inconsistent results.

\section{Shock-cloud simulations}\label{s:shkcloud}

Having verified and validated our turbulence model implementation with idealized tests, we now explore a complex problem: the astrophysical shock-cloud interaction. We solve Eqs. \ref{eq:hydro}-\ref{eq:dxieq} in \textsc{Athena} using the directionally unsplit CTU integrator \citep{1990JCoPh..87..171C} with third order reconstruction in the characteristic variables \citep{1984JCoPh..54..174C} and the HLLC Riemann solver \citep{Toro2009}. Simulations are performed on Cartesian grids in three dimensions. We use an adiabatic equation of state with the ratio of specific heats $\gamma = C_{\rm p}/C_{\rm V} = 5/3$. Self-gravity and magnetic fields are not included.

\subsection{Setup and initial conditions}\label{ss:setup}
Our simulation is a variant of the typical shock-cloud interaction: a planar shock wave of hot diffuse gas propagates through a uniform medium and impacts a cold, dense cloud. The initial conditions are determined by the Mach number of the shock $M$, the radius of the cloud $R$, and the density ratio of cloud to the ambient medium $\chi$. Our fiducial simulation uses $M = 10$, $R = 1$, and $\chi = 10$.

The ambient medium is initially uniform with density $\rho_0 = 1$ and pressure $P_0 = 1$, in arbitrary (computational) units. Our simulation domain initially extends from $-5 \le x \le 15$, $-5 \le y \le 5$, and $-5 \le z \le 5$, again in arbitrary units. All boundaries are outflow-only, except the upstream boundary (see below). The simulation resolution is indicated by the number of cells per cloud radius $N_{\rm R}$; our fiducial simulation is $N_{\rm R} = 25$, corresponding to a resolution of $512\times256\times256$. We perform a resolution test in \S\ref{sss:restest} up to $N_{\rm R} = 200$; while $N_{\rm R} = 25$ is sufficient for most quantitative estimates, the details of the mixing are notably different for $N_R \ge 100$.

The cloud begins centred at the origin and in pressure equilibrium with the ambient medium. The cloud has a spherically-symmetric density profile given by \citep[e.g.,][]{2006ApJS..164..477N}:
\begin{equation}\label{eq:profile}
\rho(r) = \rho_0 + \frac{\rho_c -\rho_0}{1 + (\frac{r}{R})^{n}},
\end{equation} where $\rho_c = \chi\,\rho_0$ is the central density and $n$ controls the steepness of the profile. We use $n=20$ to obtain a profile similar to that of P09 but steeper than that of SSS08 (which used $n = 8$). As in SSS08, we must set an arbitrary boundary for the ``cloud,'' which we denote as $r_b$ and define where $\rho(r_b) = 1.01 \rho_0$; for $R =1$ and $n=20$, $r_b = 1.25$. To trace cloud material, a passive scalar field $C_c$ is set to unity where $r \le r_b$ and zero otherwise.

We initialize the shock with the adiabatic solutions of the Rankine-Hugoniot jump conditions for a given Mach number $M$. The upstream boundary condition maintains these quantities, resulting in a shocked wind model. The shock begins at $x = -2$ and propagates in the $+x$ direction. We use an additional passive colour field to trace the mixing of shocked material in the simulation. A shock tracer $C_s$ is initialized to unity only within the leading edge of the shock with a width of one cloud radius, i.e., $C_s = 1.0$ where $-3 < x < -2$ and zero otherwise.

The time is given in terms of the ``cloud crushing time'', $t_{cc}$, defined as \citep{1994ApJ...420..213K}
\begin{equation}
t_{cc} \equiv \frac{R}{u_s} = \frac{\chi^{1/2} R}{M a_s},
\end{equation}where $u_s$ is the shock velocity within the cloud and $a_s = \sqrt{\gamma P_0/\rho_0} = \sqrt{5/3}$ is the ambient sound speed in computational units.

We do not use any mesh refinement -- simulations are run on a single mesh of uniform spacing. \textsc{Athena} is capable of static mesh refinement (SMR), which differs from adaptive mesh refinement (AMR) in that in SMR the refinement grids are placed at the beginning of the simulation and remain fixed. We did attempt to use SMR but encountered significant issues when combined with a turbulence model. Interpolation of the conserved variables (namely energy and momentum) across coarse-fine interfaces produced small numerical errors in the primitive variables (namely pressure and velocity), which were sufficient to generate artificial vorticity that was amplified by the turbulence models. Using a single grid has the further advantage that the diffusive properties of the code remain uniform across the domain.

\subsubsection{Turbulence model initial conditions}

Following GS11, we set the initial value for $k$ relative to the internal energy as $k_0 = k_i\,e_{\rm int}$ on a cell by cell basis with $e_{\rm int} = P/(\gamma - 1)$; similarly, we set the initial value for $L$ relative to the cloud radius as $L_0 = L_i\,R$. For our fiducial simulation, we choose $k_i = 10^{-2}$ and $L_i = 10^{-2}$ everywhere to roughly match the initial conditions of GS11. We note that this differs from the approach of P09 in which the authors used different initial conditions for the shock and cloud; the effect of initial conditions will be explored in \S\ref{sss:initcond}.

\subsubsection{Co-moving grid}\label{sss:comoving}

The cloud will be accelerated and disrupted by the shocked wind, and eventually all cloud material will leave the initial simulation domain. To follow the cloud evolution for as long as possible, we implement a ``co-moving grid'' similar to the method used in SSS08. We adjust the $x$-velocity at each time step to keep our domain centred on the bulk of the cloud material. At the beginning of each integration, we compute the mass-averaged cloud velocity
\begin{equation}
\langle v_x \rangle = \frac{\int_V (\rho C_c)^g\,v_x\,dV}{\int_V (\rho C_c)^g\,dV},
\end{equation}where $g$ is a weighting factor we introduce to keep the grid fixed on the densest cloud material. While SSS08 used $g=1$, we find we are better able to follow the cloud with $g=4$. We then subtract $\langle v_x \rangle$ from the $x$-velocity everywhere in the simulation and update the grid location and inflow conditions accordingly. To prevent cloud material from encroaching on the upstream boundary, we limit the co-moving velocity when cloud material would come within a distance of $2 r_b$ from the upstream boundary. We also prohibit the inflow velocity from becoming subsonic to prevent information traveling upstream. We have verified this method by comparing to simulations performed in an elongated static grid ($-5 < x < 45$); the resulting cloud evolution is nearly indistinguishable.

\begin{figure*}
  \includegraphics[width=\textwidth]{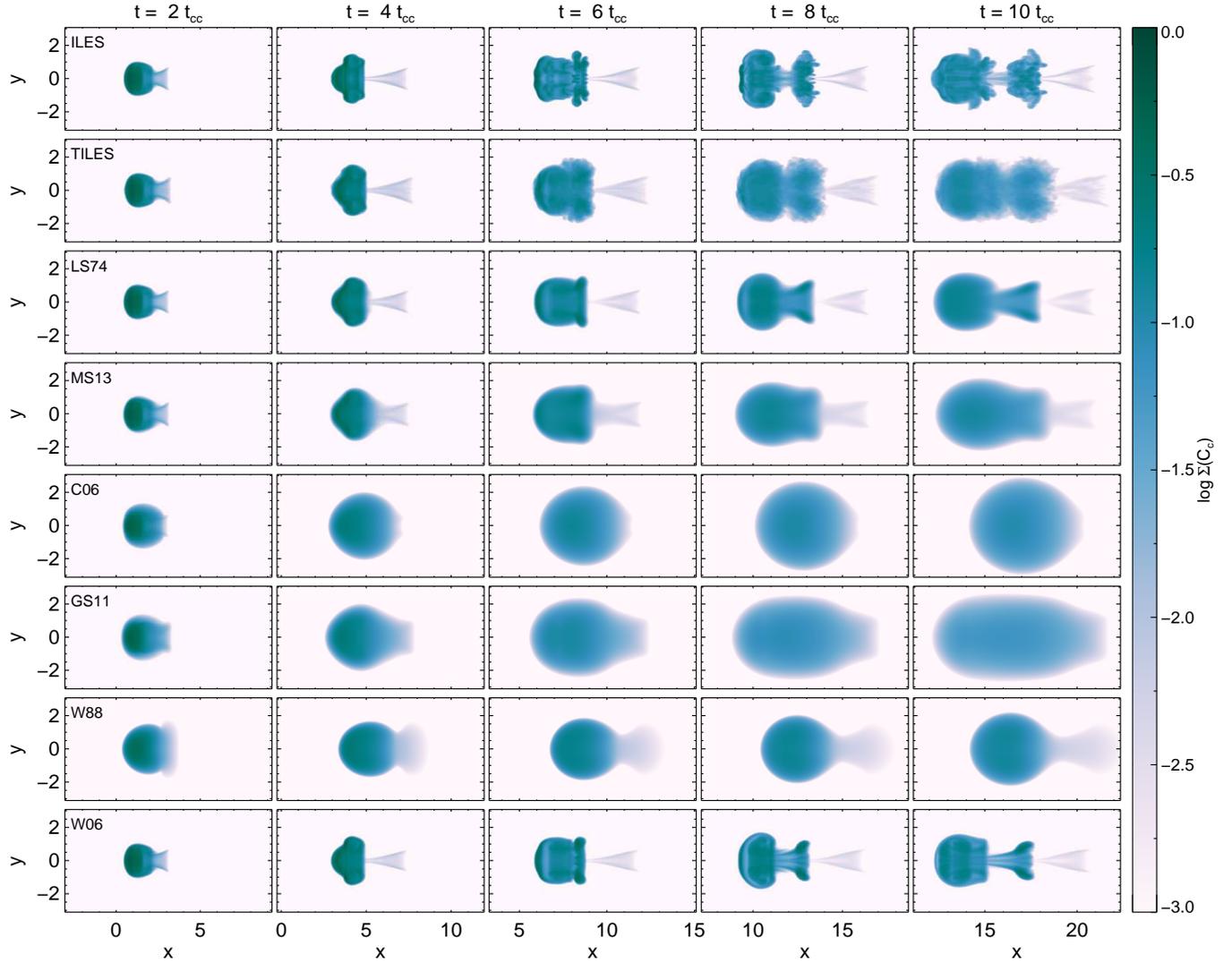}
  \caption{Time evolution of the logarithm of the density-weighted cloud column density $\Sigma (C_c) \equiv \int \rho C_c\,dz/ \int \rho\,dz$ in the fiducial ($N_{\rm R} = 25$) three-dimensional shock-cloud interaction. The units are arbitrary. From left to right, the columns show snapshots at $t =$ 2, 4, 6, 8, and 10 $t_{cc}$. From top to bottom, the rows show simulations performed with no turbulence model (ILES), ensemble-averaged grid-scale turbulence (TILES), the $k$-$\varepsilon$ models of LS74 and MS13, the $k$-$L$ models of C06 and GS11, and the $k$-$\omega$ models of W88 and W06. As the cloud is accelerated by the shock, the simulation domain moves to follow the bulk of the cloud material. The cloud is ablated forming a head-tail structure, and the characteristic vortex ring is visible at $t = 4 t_{\rm cc}$. The RANS turbulence models smooth the fluctuating structures below a characteristic length scale $L$, in some cases completely diffusing the cloud.}
  \label{f:timesnaps_colr}
\end{figure*}

\begin{figure*}
  \includegraphics[width=\textwidth]{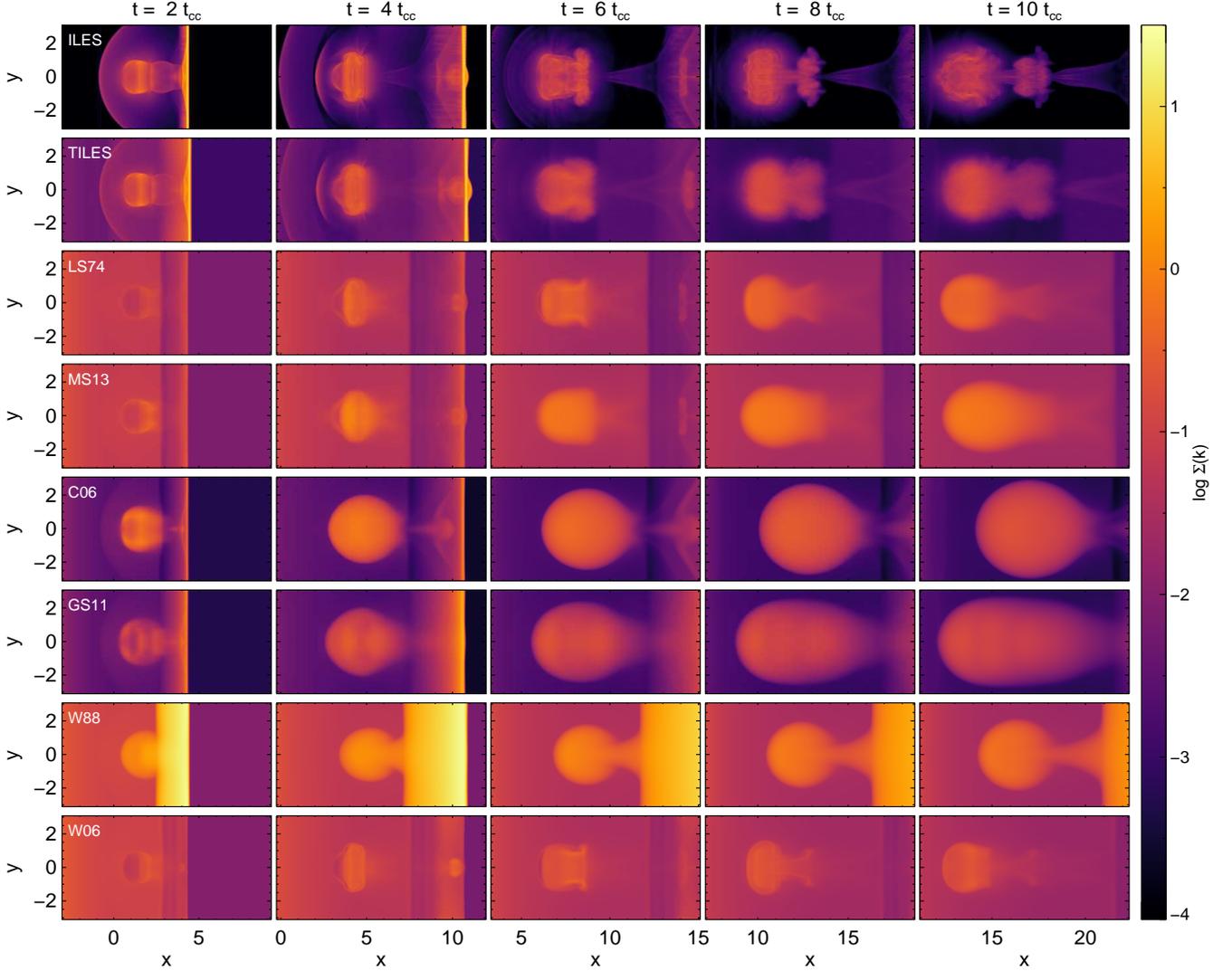}
  \caption{Similar to Figure \ref{f:timesnaps_colr}, but showing the density-weighted column of turbulent kinetic energy $\Sigma(k) \equiv \int \rho k \,dz / \int \rho \,dz $. The units are arbitrary. In runs without a turbulence model (ILES and TILES), $k$ is estimated from the resolved strain-rate tensor.}
  \label{f:timesnaps_turb}
\end{figure*}

\begin{figure*}
  \includegraphics[width=\textwidth]{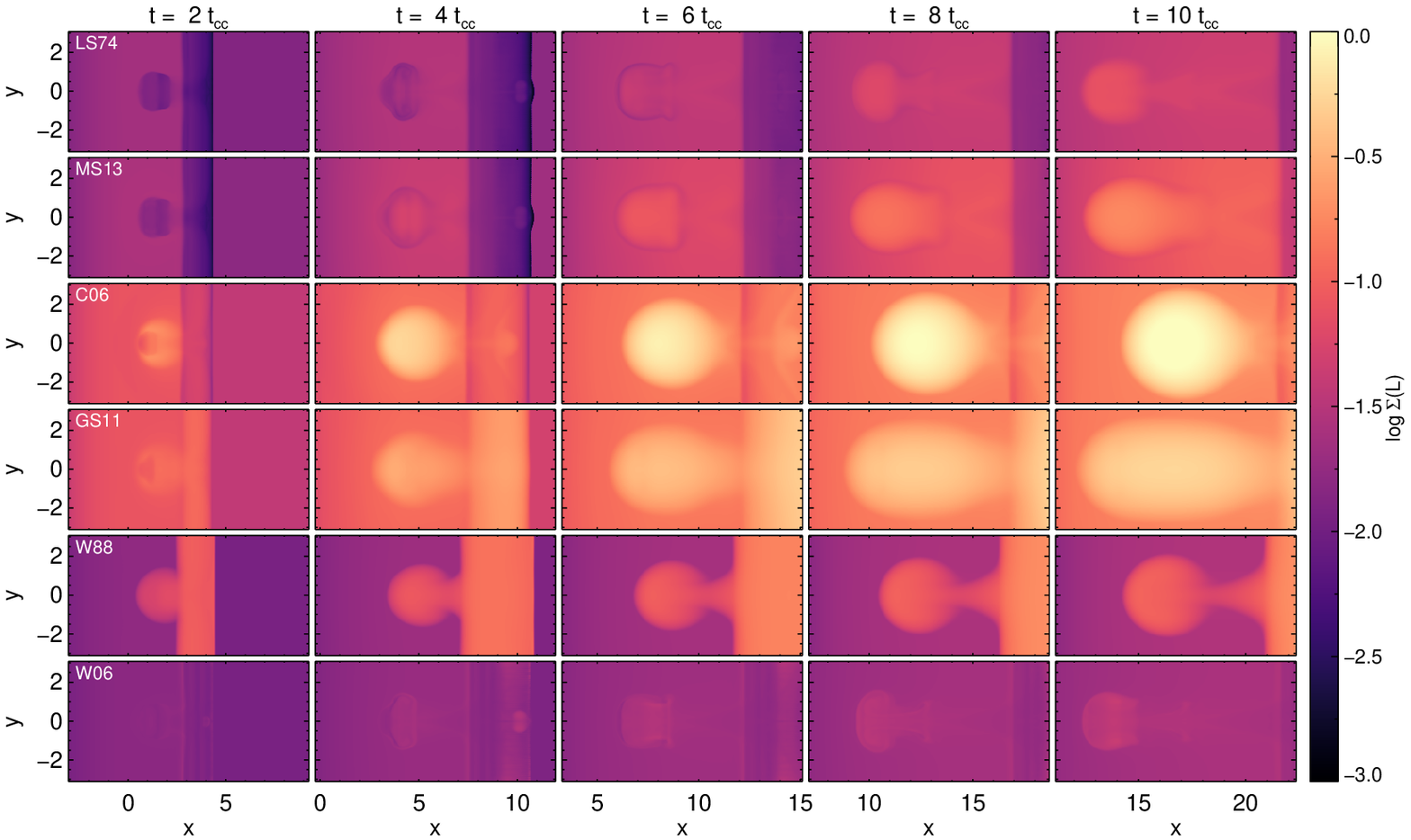}
  \caption{Similar to Figure \ref{f:timesnaps_colr}, but showing the density-weighted column of turbulent length scale $\Sigma(L) \equiv \int \rho L \,dz / \int \rho \,dz $. The units are arbitrary. The runs without a turbulence model (ILES and TILES) are not shown, as $L = \Delta$ by definition.}
  \label{f:timesnaps_tlen}
\end{figure*}

\subsubsection{Implicit Large Eddy Simulations}\label{sss:iles}

Grid-based hydrodynamics simulations performed without a turbulence model are sometimes referred to as ``inviscid'' simulations; however, the discretization of the Euler equations introduces numerical viscosity, and the turbulent cascade is truncated at the grid scale. The grid thus serves as an ``implicit'' filter, and such a simulation may be referred to as an ``Implicit Large Eddy Simulation'', or ILES \citep{Garnier2009,Schmidt2014b}. We therefore denote simulations performed without a turbulence model as ILES. We perform high-resolution ILES simulations up to $N_{\rm R} = 200$ for comparison to simulations with a turbulence model.

\subsubsection{Ensemble-averaged simulations with grid-scale turbulence}\label{sss:ensavg}

Even at high resolution, an ILES simulation with static initial conditions is not equivalent to models with a turbulence model because the turbulence models are initialized with non-zero small-scale turbulent energy ($k_0 \neq 0$). P09 therefore compared shock-cloud simulations performed with the LS74 $k$-$\varepsilon$ model to an inviscid simulation with random perturbations to the density, velocity, and pressure in the post-shock flow. We extend the P09 approach by averaging multiple high-resolution inviscid simulations initialized with different random perturbations. This should provide a good comparison, as the results from a RANS turbulence model can be interpreted as an ensemble average over many turbulent flow realizations. The velocity perturbations are drawn from a Gaussian distribution, and the width of the Gaussian is set to match the initial level of turbulence in the models, namely $k_i = 10^{-2}\,e_{\rm int}$. The amplitude of the density perturbations is drawn from a Gaussian with a width of 0.01. Note that, unlike P09, we do not perturb the pressure. We perform 10 simulations at $N_R = 25$ with different turbulent realizations and then average on a cell-by-cell basis. We refer to results from this method as ``Turbulent ILES'', or TILES.

\subsection{Diagnostics}

For comparison to previous shock-cloud simulations, we compute several standard integrated diagnostic quantities \citep{1994ApJ...420..213K}. The cloud-mass-weighted average of a quantity $f$ is defined as
\begin{equation}
\langle f \rangle = \frac{1}{M_{cl}} \int_V \rho C_c f dV,
\end{equation}where the initial cloud mass $M_{cl} = \int_{t=0} (\rho C_c)\,dV$.

We follow the effective radius normal to the $x$-axis
\begin{equation}
a = [ 5 ( \langle x^2 \rangle - \langle x \rangle ^2)]^{1/2},
\end{equation} with similar expressions along the $y$ and $z$ axes denoted $b$ and $c$ respectively. We also compute the rms velocity along each axis \citep{2006ApJS..164..477N},
\begin{equation}
\delta v_x = ( \langle v_x^2 \rangle - \langle v_x \rangle ^2 )^{1/2},
\end{equation} again with similar expressions in $y$ and $z$.

To follow the mixing, we adopt the mixing fraction $f_{\rm mix}$ introduced in \citet{1995ApJ...454..172X} and used in SSS08, where
\begin{equation}
f_{\rm mix} = \frac{1}{M_{cl}} \int_{0.1 < C_c < 0.9} \rho C_c dV.
\end{equation}As the cloud material (initially $C_c = 1.0$) is mixed into the ambient medium (initially $C_c = 0.0$), the cloud concentration will take on intermediate values and $f_{\rm mix}$ will increase.

We also examine another quantitative estimate of the mixing: the injection efficiency $f_{\rm inj}$, defined as
\begin{equation}
f_{\rm inj} = \frac{1}{\eta M_{s}} \int_{C_c \ge 0.1} \rho C_s dV,
\end{equation} where $M_s = \int_{t = 0} \rho C_s dV$ is the initial shock tracer mass and $\eta$ is a normalization factor. As the shock passes over the cloud, mixing at the leading edge by RT instabilities and at the edges by KH instabilities will ``inject'' shock material into the cloud. This is of particular interest for studies of chemical enrichment of the early Solar system with short-lived isotopes from supernovae \citep{2016MNRAS.462.2777G}. The injection efficiency is normalized via $\eta$ such that only the mass of the shock tracer directly incident on the cloud cross-section $\pi r_b^2$ is considered; hence, $f_{\rm inj} = 1$ indicates ``perfect'' injection.

\subsection{Results}\label{ss:results}

\subsubsection{Dynamical evolution}

We follow the interaction of the shocked wind with the cloud for up to 10 cloud-crushing times. Figure \ref{f:timesnaps_colr} shows the time evolution of the cloud column density $\Sigma (C_c) = \int \rho C_c\,dz / \int \rho\,dz$ in the fiducial ($N_{\rm R} = 25$) simulations for each of the models, including no turbulence model (ILES) and ensemble-averaged grid-scale turbulence (TILES). The cloud material is initially confined within $r \le r_b$, but after impact material is ablated and mixed into the shock and ambient medium, leading to a head-tail structure. The cloud is accelerated in the $+x$-direction; as described in \S\ref{sss:comoving}, we shift our grid to be co-moving with the densest cloud material. The location of the cloud at a given time varies from run to run, as each turbulence model uniquely alters the cloud acceleration and destruction. As material is ablated from the edges of the cloud, large KH rolls develop in the ILES simulation. Around 4~$t_{cc}$, the characteristic vortex ring is clearly evident. The evolution of an inviscid adiabatic shock-cloud interaction is described in detail in PP16; we here focus on the differences resulting from the turbulence models.

The turbulence models also include diffusion of passive colour fields, which is of particular importance for the mixing estimates. In the ILES simulations, cloud material is most concentrated at the cloud edges as a result of the KH instability. The additional viscosity from the turbulence models diffuses the colour field to varying degrees. In the models of LS74, MS13, and W06, three structures still remain in the colour field: the dense head, the vortex ring, and the diffuse tail. However, in the models of C06, GS11, and W88, the colour field is largely smoothed. In C06 and GS11, the cloud material becomes nearly uniformly distributed in an oblate spheroid. It is unclear whether this is due to increased buoyancy, shear effects, and/or over-production of turbulent energy.

Figure \ref{f:timesnaps_turb} presents the time evolution of the density-weighted column of specific turbulent energy $\Sigma (k) = \int \rho k \,dz / \int \rho\,dz$. For the ILES and TILES runs, the turbulent energy is not explicitly tracked; we therefore follow \citet{2011A&A...528A.106S} and construct an estimate for $k = C_k \Delta^2 |S^*|^2$, where $\Delta$ is the grid resolution, $S^*_{ij} = S_{ij}-(1/3)\delta_{ij}S_{kk}$ is the trace-free resolved strain rate tensor (see Eq. \ref{eq:sij}), and $C_k$ is a scaling constant. The exact scaling is uncertain; \citet{2011A&A...528A.106S} used $C_k \approx 0.013$ based on supersonic isothermal turbulence. Here, we set $C_k = 1$ and treat $k$ as a morphological rather than quantitative estimate.

Figure \ref{f:timesnaps_turb} shows that in all runs the strongest areas of turbulence generation are 1) at the cloud edges due to shearing motions; 2) in the cloud tail due to shear and compression; and 3) at the shock front due to compression. LS74 and W06 produce relatively little turbulence, resulting in a correspondingly low turbulent viscosity. These models produce only slight differences in morphology from the ILES and TILES cases. While the small-scale structure is smoothed, the two large KH rolls are still present. In contrast, W88 produces large amounts of turbulent energy, particularly in the shock. The turbulent pressure term ultimately leads to non-physical spreading of the shock downstream. The strong shear at the cloud edges spreads material into two primary streamers. This also occurs in MS13, but the dominant turbulence is at the leading edge of the cloud due to the inclusion of buoyancy effects (RT instability). A similar effect is seen in C06 and GS11 due to the buoyancy; however, in C06 and GS11 the ambient turbulence dissipates rapidly and the cloud expands due to the increased interior turbulent pressure.

The transmitted shock within the cloud also increases the turbulent length scale $L$ via the dilatation term ($\nabla \cdot \bf{u}$) in $\Psi_L$ of the $k$-$L$ models (C06 and GS11); this is seen in Figure \ref{f:timesnaps_tlen}, which shows the evolution of the density-weighted column of $L$, $\Sigma (L) = \int \rho L\,dz / \int \rho\,dz$. These models show turbulent length scales roughly an order of magnitude greater than the other models, while the turbulent kinetic energy is roughly an order of magnitude lower. The most similar model is MS13; however, all three models with buoyancy terms (MS13, C06, and GS11) show significant expansion, and the cloud is eventually diffused completely. The models without a turbulence model (ILES and TILES) are not shown in Figure \ref{f:timesnaps_tlen}, as $L$ would simply be the grid scale $\Delta$.

\subsubsection{Evolution of diagnostic quantities}\label{sss:evodiag}

\begin{figure}
  \includegraphics[width=\columnwidth]{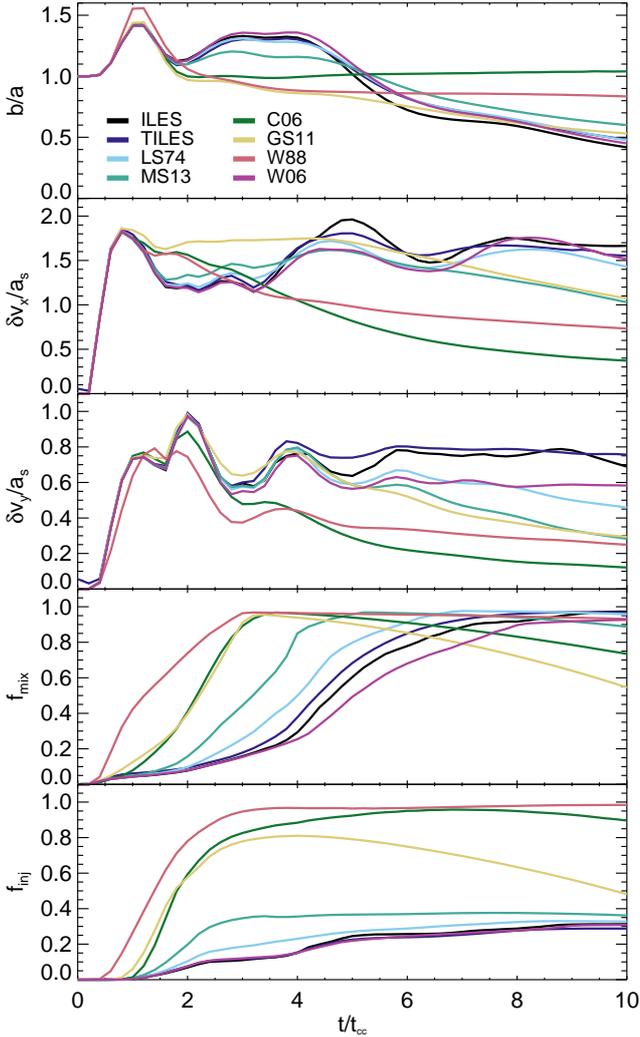}
  \caption{Evolution of diagnostic quantities in the three-dimensional shock cloud simulation. From top to bottom, the cloud axis ratio $b/a$; the rms velocity along the $x$-axis relative to the sound speed $a_s$; the rms velocity along the $y$-axis relative to the sound speed $a_s$; the mixing fraction $f_{\rm mix}$; and the injection efficiency $f_{\rm inj}$. The turbulence model is indicated by line colour, with the inviscid case shown in black. The units are arbitrary.}
  \label{f:mdlcompare}
\end{figure}

Figure \ref{f:mdlcompare} shows the time evolution of various diagnostic quantities. Overall, the turbulence models produce similar results for the cloud axis ratio $b/a$, excepting C06 and W88. In C06, large amounts of turbulent pressure within the cloud cause the cloud to expand and become spherical. However, the turbulence models show little agreement in their treatment of either motions ($\delta v$) or mixing ($f_{\rm mix}$ and $f_{\rm inj}$). The ILES and TILES simulations are comparable, but all simulations with a turbulence model show reduced rms velocity dispersions, as the additional turbulent viscosity diffuses the small-scale turbulent motions. Recall that the turbulence models work by averaging out the fluctuating velocities below the characteristic length scale. C06 and GS11 lead to the largest values of $L$ -- on the order of the cloud radius within the cloud -- and therefore smooth nearly all small-scale fluctuations.

This also affects the mixing. The TILES model shows only slightly faster mixing than the ILES result. This differs from what was observed by P09, where the mixing of material proceeded almost twice as fast in models with grid-scale turbulence compared to those without (see, e.g. fig. 15g of P09, where $m_{\rm core}$ is an alternative measure for mixing). This is mostly likely due to the strength of the imposed turbulence, which was considerably higher in P09 than in our TILES simulations.

As already noted, LS74 and W06 introduce the least turbulent viscosity and therefore most resemble the ILES case. Surprisingly, W06 shows a reduction in $f_{\rm mix}$ relative to the ILES runs. In all runs, $f_{\rm mix}$ approaches unity, indicating complete cloud disruption. In several models, the expansion of the cloud at late times reduces the concentration of cloud colour field below the mixing threshold ($C_c \ge 0.1$) which causes $f_{\rm mix}$ to decrease. A different trend is observed in the injection efficiency, where the three most diffusive models (W88, C06, and GS11) reach a significantly different peak value from the other models. Both the shock and cloud are diffused, and the increased viscosity leads to enhanced injection. There is agreement between most models at a final value of $f_{\rm inj} \approx 0.3$ -- slightly higher than previous shock-cloud studies of Solar system enrichment, which found $f_{\rm inj} \lesssim 0.1$ \citep[e.g.][]{2015ApJ...809..103B,2016MNRAS.462.2777G}.

\subsubsection{Model validity}\label{sss:mdlvalid}

\begin{figure}
  \includegraphics[width=\columnwidth]{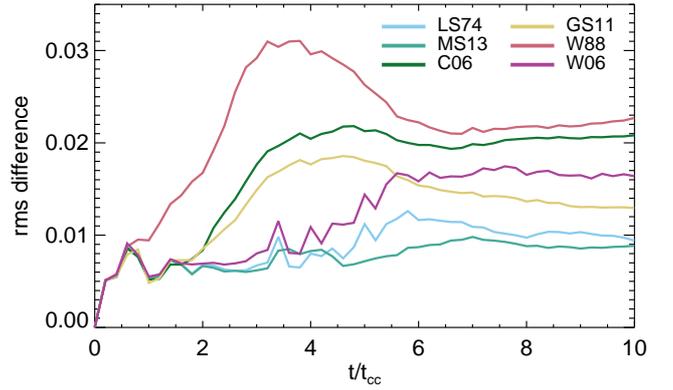}
  \caption{Time evolution of the rms difference between the TILES result and the turbulence models for the density-weighted cloud colour field $\Sigma(C_c)$. The turbulence model is indicated by line colour. LS74 and MS13 show the best agreement with the TILES result.}
  \label{f:mdldiffplot}
\end{figure}

A primary goal of this work is to compare the behavior of turbulence models in an identical astrophysical application. Clearly the models do not all reproduce the same dynamical and quantitative evolution. As noted in Section \ref{sss:ensavg}, we believe the best reference for a RANS model is an ensemble-average of high-resolution grid-scale turbulence simulations. We therefore compare the turbulence model results to the TILES result. We compute an rms difference for the density-weighted cloud colour field at each time step using the TILES result as the reference. The time evolution of the rms difference is shown Figure \ref{f:mdldiffplot}. We observe that the $k$-$\varepsilon$ models of LS74 and MS13 agree best with the TILES result. A similar trend is observed when compared to the highest resolution ILES simulation ($N_R = 200$, see \S\ref{sss:restest}).

\subsubsection{Effect of compressibility corrections}\label{sss:cceff}

As seen in \S\ref{ss:mixcompcorr}, the RANS models here considered are largely calibrated with subsonic, incompressible experiments, and they do not reproduce the correct shear layer growth rate without modifications. As our shock is supersonic ($M = 10$), we anticipated a compressibility correction would be important to model the evolution. However, we find that the compressibility corrections have a negligible effect on the simulation evolution in LS74, MS13, W88, and W06. We do not test GS11 without $\tau_{\rm KH}$, as this could affect the calibration; and we do not test C06, as there is no straightforward way to implement a correction. As the results are nearly indistinguishable, we do not present any figures. It is possible that the effects may become important at higher Mach numbers, but we defer this for future studies.

\subsubsection{Dependence on initial conditions}\label{sss:initcond}

\begin{figure*}
  \includegraphics[width=\textwidth]{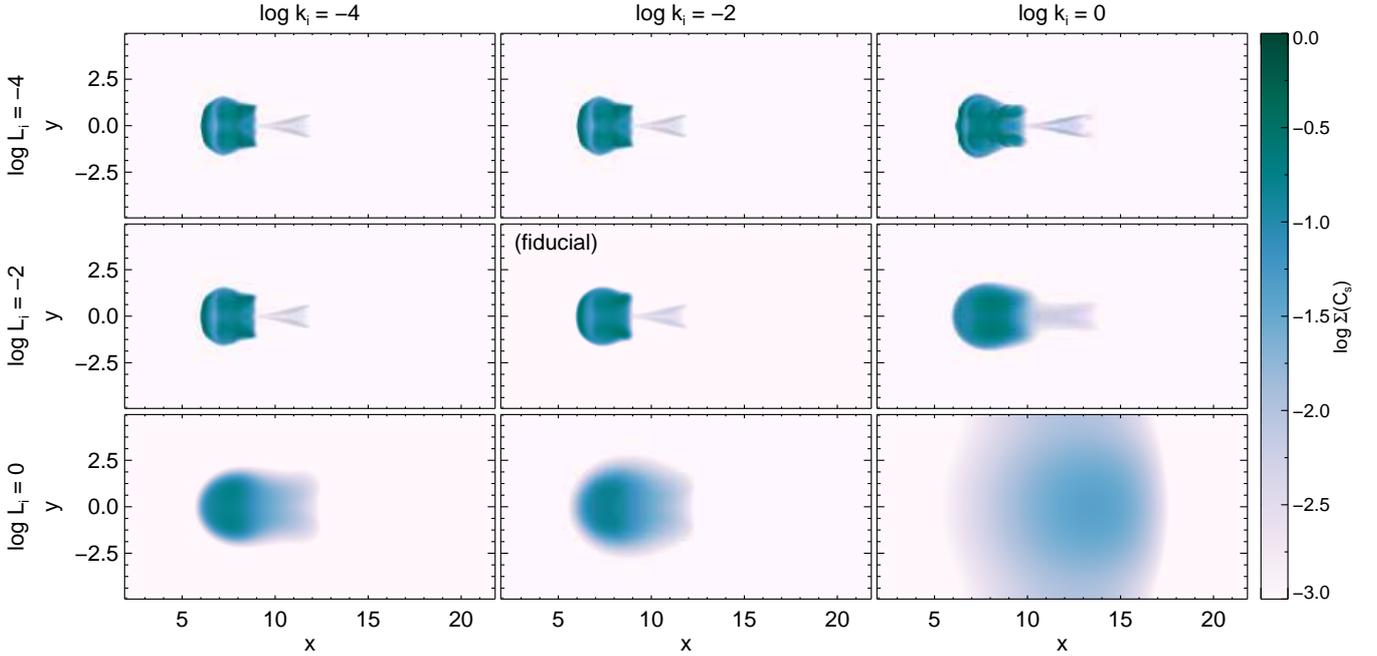}
  \caption{Snapshots of the density-weighted cloud column density $\Sigma (C_c)$ at $t=6 t_{\rm cc}$ for different initial conditions with the LS74 model. The columns show varying levels of initial turbulent energy $k_0 = k_i e_{\rm int}$; from left to right, $k_i = 10^{-4}$, $10^{-2}$, and 1. The rows show varying initial turbulent length scale $L_0 = L_i R_c$; from top to bottom, $L_i = 10^{-4}$, $10^{-2}$, and 1. Increasing either quantity increases the turbulent viscosity and hence the diffusion of the colour field. The units are arbitrary; $\epsilon_0$ can be determined for each model using the relation in Section \ref{ss:keps}.}
  \label{f:ictest_colr}
\end{figure*}

\begin{figure}
  \includegraphics[width=\columnwidth]{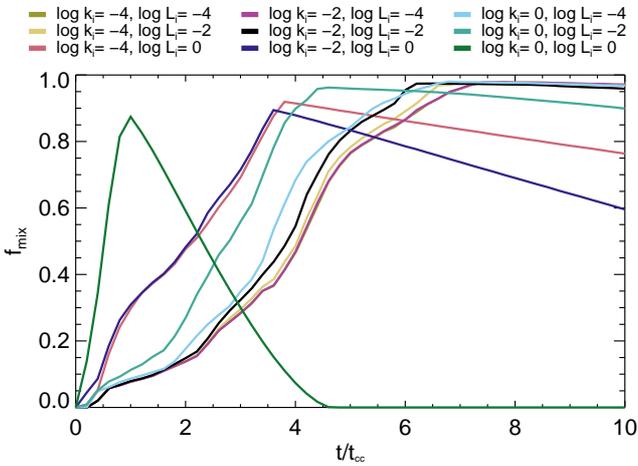}
  \caption{Time evolution of the mixing fraction $f_{\rm mix}$ for different initial conditions with the LS74 model. The colours show different combinations of initial turbulent energy $k_i$ and initial turbulent length scale $L_i$. As the cloud is diffused, the cloud concentration $C_c$ drops below the threshold and $f_{\rm mix}$ decreases.}
  \label{f:icplot}
\end{figure}

The RANS turbulence models considered here are known to be sensitive to initial conditions, particularly the W88 model \citep{Wilcox2008}. In most astrophysical applications, the prescription for the initial values of $k$ and $L$ is arbitrary. We set the initial value for $k$ relative to the internal energy as $k_0 = k_i\,e_{\rm int}$ and for $L$ relative to the cloud radius as $L_0 = L_i\,R$. Our fiducial simulation uses $k_i = 10^{-2}$ and $L_i = 10^{-2}$ to roughly match the initial conditions of GS11. However, P09 chose non-uniform initial conditions, with varying levels of $k$ between the shock and the cloud. Similar to P09, we test the dependence of the LS74 turbulence model on the initial conditions by performing simulations with varying levels of initial turbulence $k_i$ and length scale $L_i$, ranging from $10^{-4}$ to $10^{0}$ in both quantities. We perform this test at $N_R = 12$, as the increased viscosity decreases the allowed time step size.

Figure \ref{f:ictest_colr} presents a snapshot of the density-weighted average cloud colour column at $t=6  t_{\rm cc}$ for each combination of $k_i$ and $L_i$ in the LS74 model. We see that even an order of magnitude difference in either quantity produces notable differences in the evolution and mixing. Increasing either $k$ or $L$ increases the turbulent viscosity, to the point where the cloud is completely diffused into the background. This is also evident in Figure \ref{f:icplot}, which shows the time evolution of the mixing fraction $f_{\rm mix}$ in runs with different initial conditions for the LS74 model. Our results agree with earlier findings by P09, in which simulations with low initial turbulence ($k_i = 10^{-6}$ in the shock) showed decreased mixing (as evidenced by e.g, a slower decrease in core mass $m_{\rm core}$ in fig. 15g of P09) compared to simulations with higher initial turbulence ($k_i = 0.13$ in the shock). It is perhaps not surprising that different initial conditions produce different results, as each represents a particular physical state (i.e., more or less turbulence at varying scales). One should carefully consider the initial conditions when using RANS models in an unsteady flow.

Finally, PP16 concluded that the LS74 $k$-$\varepsilon$ model did not significantly affect the evolution of their three-dimensional shock-cloud simulations. However, this is most likely due to their choice of initial conditions; PP16 used $k_i = 10^{-6}$ and $L_i = 1.6\times10^{-4}$ (Pittard, personal communication) in all simulations, corresponding to very low initial levels of turbulence. While the LS74 model has very little effect for small (and probably reasonable) initial values of $k$ and $L$, we demonstrate that the model can dramatically alter 3D simulations under certain conditions.

\subsubsection{Resolution dependence}\label{sss:restest}

\begin{figure}
  \includegraphics[width=\columnwidth]{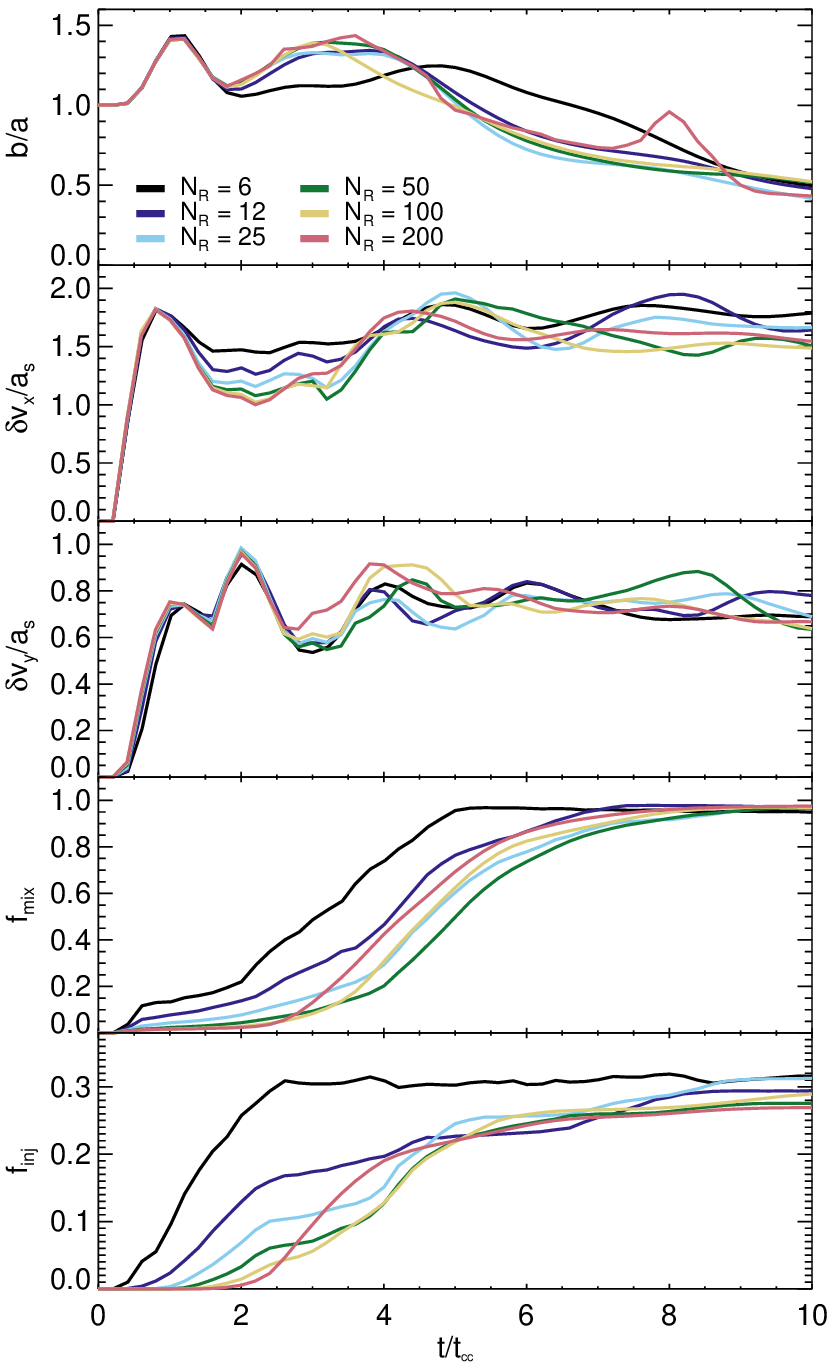}
  \caption{Similar to Figure \ref{f:mdlcompare}, but for different simulation resolutions (measured in cells per cloud radius $N_R$) with no turbulence model (ILES). We observe only small variance in the evolution of global quantities (axis ratio $b/a$ and velocity dispersions $\delta v$) for $N_R \gtrsim 25$. The mixing estimates ($f_{\rm mix}$ and $f_{\rm inj}$) decrease with increasing resolution up to $N_{\rm R} = 50$, but then increase again with increasing resolution.}
  \label{f:resplot}
\end{figure}

\begin{figure}
  \includegraphics[width=\columnwidth]{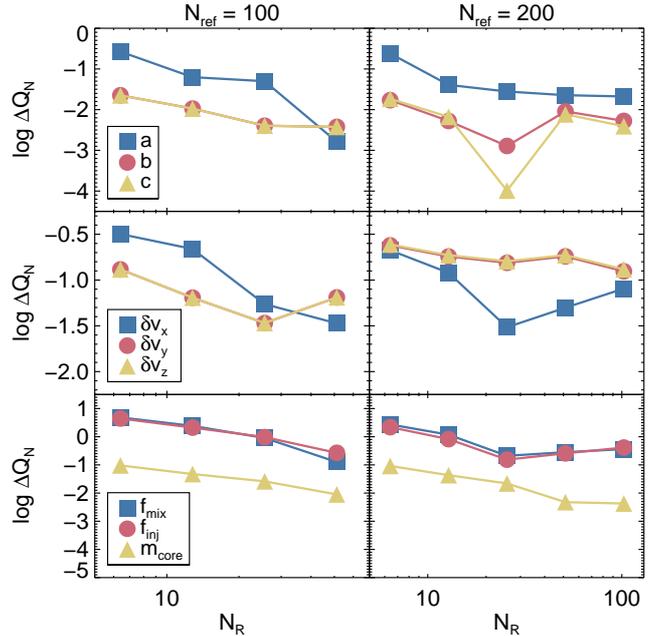}
  \caption{Estimates of the relative difference $\Delta Q_N$ as a function of resolution for global quantities (top row: effective radii; middle row: rms velocities) and mixing estimates (bottom row: mixing, injection, and core mass) at $t = 3 t_{\rm cc}$ in the ILES simulations. The left column uses $N_R = 100$ as the reference resolution, while the right column uses $N_R = 200$. As in fig. A13 of PP16, we see decreasing relative difference with increasing resolution when $N_R = 100$ is used as the reference, which appears to indicate convergence. However, when $N_R =200$ is used as the reference, there is no obvious sign of convergence. We attribute this to the partial resolution of the turbulent cascade for $N_R \gtrsim 50$.}
  \label{f:relerrplot}
\end{figure}

\begin{figure}
  \includegraphics[width=\columnwidth]{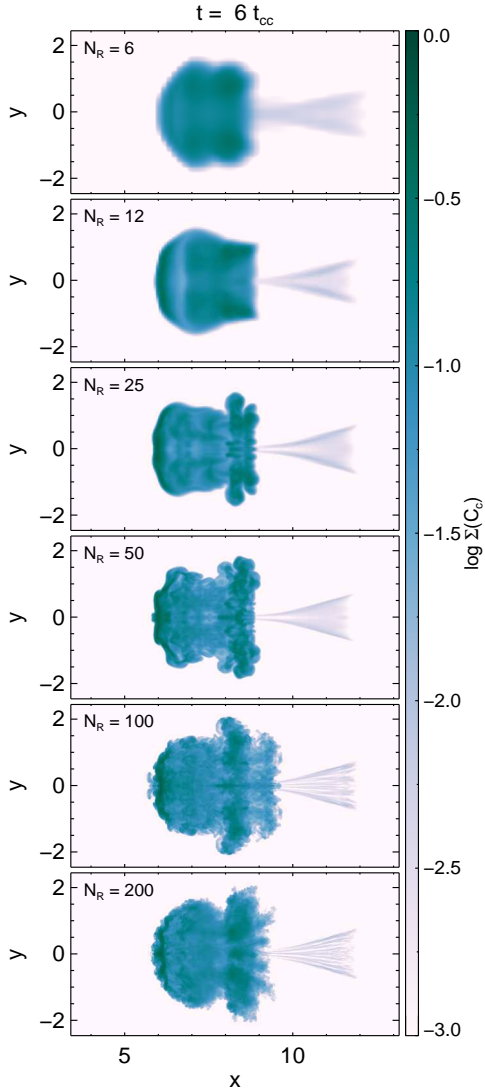}
  \caption{Snapshots of the density-weighted cloud column density $\Sigma (C_c)$ at $t= 6 t_{\rm cc}$ for different simulation resolution with the inviscid (ILES) model. The resolution increases from top to bottom, from $N_R = 6$ up to $N_R = 200$. Above $N_R = 50$, the reduced numerical viscosity allows the growth of KH instabilities and enhances the mixing. The units are arbitrary.}
  \label{f:snapshot}
\end{figure}

\begin{figure}
  \includegraphics[width=\columnwidth]{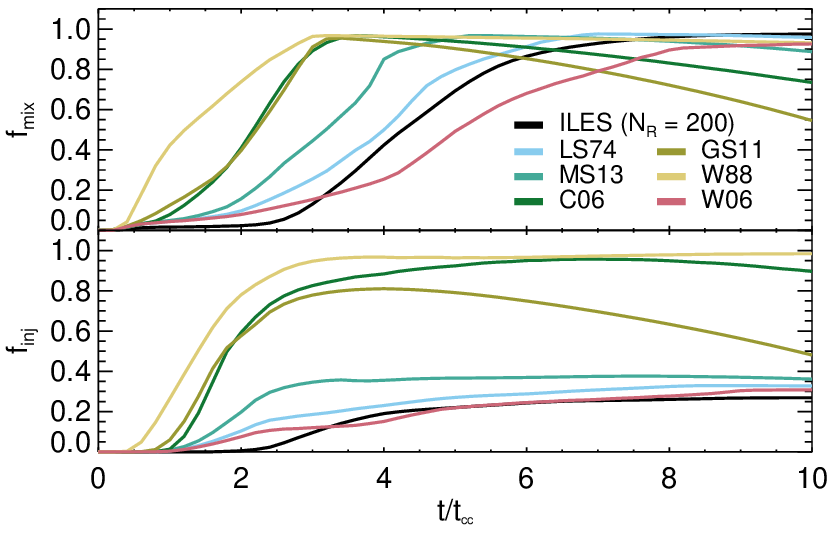}
  \caption{Similar to Figure \ref{f:mdlcompare}, but only the mixing estimates are shown for the highest resolution ILES simulation ($N_R = 200$) and the turbulence models ($N_R = 25$). All models other than W06 show increased mixing relative to the ILES result, and LS74 and W06 show the best agreement.}
  \label{f:mdlmixcompare}
\end{figure}

While 100 cells per cloud radius are necessary to see convergence of global quantities in 2D studies \citep[][P09]{1994ApJ...420..213K}, the resolution limit may be less strict in 3D. PP16 found that 32--64 cells may be sufficient for global convergence in 3D simulations. Figure~\ref{f:resplot} shows the time evolution of the diagnostic quantities in ILES simulations for resolutions $N_R = 10-200$. In agreement with PP16, we observe that globally-averaged quantities ($b/a$ and $\delta v$) exhibit only small variation with increasing resolution for $N_R \gtrsim 25$.

However, it is difficult to assess whether or not this represents true convergence. For consistency with previous work, we perform an analysis similar to that described in Appendix A3 of PP16. We calculate the relative difference $\Delta Q_N$ between a measurement $Q$ at a given resolution $N$ and the same measure at a reference resolution $N_{\rm ref}$ (typically the highest resolution), given by eq. A1 of PP16 as
\begin{equation}
\Delta Q_N = \frac{|Q_N - Q_{N_{\rm ref}}|}{|Q_{N_{\rm ref}}|}.
\end{equation} Figure \ref{f:relerrplot} shows the relative difference as a function of simulation resolution $N_R$ for various diagnostic quantities at $t=3 t_{\rm cc}$. We compare results using $N_{\rm ref} = 100$ (as in PP16) and $N_{\rm ref} = 200$. We note that our axial direction is $x$, whereas in PP16 the axial direction is $z$; hence our quantity $a$ should be compared to $c$ in e.g., fig. A13 of PP16, and likewise our $\delta v_x$ to their $\delta v_z$. For further comparison with PP16, we also calculate $\Delta Q_N$ for the core mass, $m_{\rm core}$, defined as
\begin{equation}
m_{\rm core} = \int_{C_c \ge 0.5} \rho C_c dV.
\end{equation}We finally note that our initialization of the cloud colour field is slightly different than in PP16; we use a constant value of $C_c = 1$ for $r \le r_b$, while PP16 used a spatially varying $C_c$ that decreased with increasing radius within the cloud.

If we use $N_R = 100$ as our reference resolution (left column of Figure \ref{f:relerrplot}), we find good agreement with PP16. The relative difference decreases with increasing resolution for most quantities, suggesting convergence. The only quantities with increasing difference are the velocity dispersions along axes perpendicular to the flow ($\delta v_y$ and $\delta v_z$), which are not shown in fig A13 of PP16. However, the trend is less certain if we use our highest resolution simulation with $N_R = 200$ as the reference. There is no longer any sign of convergence, particularly in the mixing measures.

This is surprising given previous studies of the shock-cloud interaction. \citet{1995ApJ...454..172X} found little variance in $f_{\rm mix}$ up to $N_R = 50$ in hydrodynamical shock-cloud interactions. While similar magneto-hydrodynamical simulations by SSS08 did not show convergence in $f_{\rm mix}$ up to $N_R \approx 120$, the authors predicted that, in simulations without an explicit viscosity, $f_{\rm mix}$ should continue to decrease with increasing resolution and tend to zero at infinite resolution. In examining the time evolution in Figure \ref{f:resplot}, we do not observe either trend. While we find that $f_{\rm mix}$ does show a decreasing trend up to $N_R = 50$, $f_{\rm mix}$ actually increases with increasing resolution beyond this point. A similar result is observed in fig. A8a of PP16; the mixing (as measured by $m_{\rm core}$) decreases with increasing resolution up to $N_R = 64$, at which point increased mixing (indicated by a faster decrease in $m_{\rm core}$) is observed for $N_R = 128$.

These results suggest that for resolutions $N_R \gtrsim 50$, mixing in the ``inviscid'' hydrodynamical shock-cloud simulation starts to be dominated by turbulent diffusion rather than numerical diffusion. If the correlation time of the turbulence is short compared to the numerical diffusion time, the turbulent viscosity will dominate the diffusion \citep[see the Appendix of][]{2004ApJ...603..165H}. At low resolutions ($N_R \le 50$), the numerical viscosity dominates the dynamics and affects the growth of instabilities. As the resolution increases up to $N_R=50$, numerical diffusion decreases, yet the turbulent cascade is not yet sufficiently resolved to show ``true'' turbulent mixing, i.e., mixing rates independent of the numerical diffusion. The mixing is therefore at a minimum near this resolution, which could explain the apparent ``convergence'' observed in Figure \ref{f:relerrplot} when $N_R = 100$ is used as the reference.

For $N_R \gtrsim 50$, the numerical viscosity is reduced to the point that the RT and KH instabilities can grow at the cloud surface and seed further turbulent motions. This is evident in Figure \ref{f:snapshot}, which shows a snapshot of the cloud column density at $t=6 t_{\rm cc}$ for varying simulation resolution. At high resolution, the leading edge of the cloud is saturated with RT fingers, and the shear at the cloud edge generates KH rolls that spawn additional vortices in the cloud wake. The turbulent cascade that develops is now largely resolved; the corresponding Reynolds number is large, and the mixing is increased.

The continued increase in mixing from $N_R = 100$ to $N_R = 200$ in our fiducial simulation suggests that the turbulent cascade is still not fully resolved at this point. It is unclear whether the mixing would continue to increase with increasing resolution. As our simulations are performed on a fixed grid with no mesh refinement, extending our simulations beyond $N_R = 200$ is not feasible given the computational burden (see Appendix).

We are also unable to perform simulations with $N_R > 25$ when using a turbulence model, due to the stability requirement that $dt \lesssim (\Delta)^2$. P09 found that the LS74 model reduced the convergence requirements in 2D, but PP16 found the model had little effect in 3D. As noted in Section \ref{sss:initcond}, this may be a consequence of the low level of initial turbulence used in PP16. In our resolution tests up to $N_R = 25$, we find no significant benefit from the turbulence models.

Figure \ref{f:mdlmixcompare} compares the time evolution of the mixing estimates for the ILES model at $N_R = 200$ with the turbulence models at $N_R =25$. Despite the increased mixing at $N_R = 200$, all turbulence models other than W06 still indicate more mixing than observed. Yet if the ILES mixing continues to increase at higher resolutions, as the trend suggests, it may be that the turbulence models effectively predict the ``correct'' mixing.

\subsubsection{Dependence on numerical methods}\label{sss:methodtest}

\begin{figure}
  \includegraphics[width=\columnwidth]{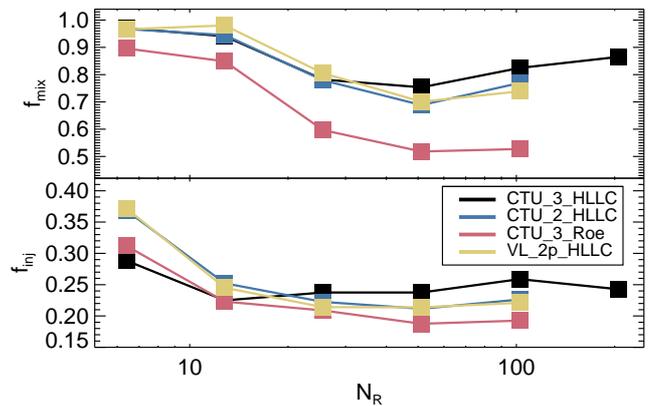}
  \caption{Estimates of the mixing fraction $f_{\rm mix}$ and injection efficiency $f_{\rm inj}$ at $t=6 t_{\rm cc}$ as a function of resolution, indicated by the number of cells per cloud radius $N_R$. Results are shown for different algorithms in the Godunov scheme; we test the effect of the integrator (CTU, VL), the order of spatial reconstruction (3, 2, 2p), and the Riemann solver (HLLC, Roe). The Riemann solver has the greatest effect, reducing the mixing estimates and failing to show an increase in mixing for $N_R > 50$.}
  \label{f:sgtmresplot}
\end{figure}

Figure \ref{f:sgtmresplot} shows the resolution dependence of the mixing estimates at $t = 6 t_{\rm cc}$ for various combinations of integrators, Riemman solvers, and reconstruction accuracy. Our fiducial simulation uses the CTU integrator with 3rd order reconstruction of the characteristic variables and the HLLC Riemann solver (denoted CTU\_3\_HLLC). We also test second order reconstruction (CTU\_2\_HLLC); the Roe Riemann solver \citep{Roe1981} with H-correction \citep{2008ApJS..178..137S} (CTU\_3\_Roe); and the Van Leer (VL) integrator \citep{2009NewA...14..139S} with second order reconstruction in the primitive variables (VL\_2p\_HLLC). We find that changing any of these algorithms in the Godunov scheme can alter the degree of mixing, especially the Riemann solver. The results obtained with the Roe solver are almost a factor of two below the fiducial results; furthermore, it does not show the trend of increasing $f_{\rm mix}$ from $N_R = 50$ to $N_R = 100$ as seen in the other runs. The dependence of ILES mixing on the numerical algorithm underscores the utility of a turbulence model.

\section{Discussion}\label{s:discussion}

In an effort to understand previous shock-cloud simulations, we have limited our exploration to RANS turbulence models. However, LES models are probably more appropriate for most astrophysical applications, including the shock-cloud interaction. The RANS approach tends to diffuse the small scale structure in the simulation, yet these are often the scales of greatest interest in astrophysics applications (e.g., star formation). In contrast, the resolved dynamics are largely unaffected in LES, and the filtering approach is ideal for unsteady flows. Despite these differences in formulation, the methods of LES are remarkably similar to RANS; the models have similar equations with similar closures, such as eddy-viscosity and gradient-diffusivity. The simplest LES model is the Smagorinsky model \citep{1963MWRv...91...99S}, which is essentially a zero-equation mixing-length model. The LES model of \citet{2011A&A...528A.106S} is a one-equation model; $k$ is followed with a transport equation, while the turbulent length scale $L$ is simply replaced by the grid scale spacing $\Delta$. LES models also suffer the same calibration issues as RANS. \citet{2011A&A...528A.106S} calibrated their model using high-resolution ILES simulations of turbulence, but it is difficult to determine if this approach is valid (see \S\ref{sss:restest} and Figure \ref{f:snapshot}).

We have only tested two-equation models. Models with fewer equations, such as the one-equation Spalart-Allmaras model \citep{Spalart1992}, are easy to implement but do not perform well in situations with inhomogeneous or decaying turbulence. However, models with two or fewer equations make use an isotropic eddy-viscosity. This assumption of isotropy severely limits the accuracy of these models in regions of high vorticity. Anisotropic models, such as the seven-equation Reynolds-Stress-Transport model \citep{Wilcox2008}, independently follow the six components of the turbulent stress tensor plus a dissipation equation. This approach is highly accurate, but the associated computational cost is often prohibitive. One compromise may be the use of a non-linear eddy-viscosity relation, such as that used in \citet{2011A&A...528A.106S}. All of the RANS models considered here use linear eddy-viscosity relations, but the additional complexity of the non-linear relation improves results in complex flows without the need for additional stress transport equations \citep{Gatski2000}.

We also note that the assumption of isotropy is incorrect in magnetized turbulence \citep{1995ApJ...438..763G}, as typically encountered in astrophysical applications. Eddies are stretched along the field lines, and the anisotropy is scale-dependent and increases toward smaller-scales \citep{2003MNRAS.345..325C}. It is unclear if an anisotropic RANS model could be developed for magnetohydrodynamics (MHD); however, such models could be developed in the LES framework \citep{2015SSRv..194...97M}. Indeed, closures for the MHD LES equations have been proposed \citep{2016PhPl...23f2316V} but such methods have yet to be thoroughly validated.

One potential benefit of a turbulence model is the proper modeling of the RT instability \citep{Dimonte2004}. However, the buoyant turbulence models here considered seem to perform poorly in complex flows and generate excessive turbulence. Critically, the models have not been validated for use in supersonic, highly compressible turbulence, which is exactly the regime of interstellar gas dynamics. While compressibility corrections can be used, simulations have demonstrated that they are physically incorrect \citep{Vreman1996}.

Finally, we note that we are limited in our use of turbulence models by an explicit time integration method -- maintaining stability requires $dt \propto (\Delta)^2$. Implicit formulations are possible \citep[e.g.][]{HUANG1992} but the associated computational cost may be significant due to coupling between the turbulent variables.

\section{Conclusions}\label{s:conclusions}

We have developed a common framework for two-equation RANS turbulence models in the \textsc{Athena} hydrodynamics code. All models use a linear eddy-viscosity relation based on resolved dynamics to add turbulent diffusivity. We have implemented six RANS turbulence models: the $k$-$\varepsilon$ models of LS74 and MS13; the $k$-$L$ models of C06 and GS11; and the $k$-$\omega$ models of W88 and W06.

We have verified the models with the subsonic shear mixing layer. The models can only reproduce the correct mixing layer growth rate for certain definitions of the layer width $\delta$ (Figure \ref{f:chiravalle_growth}), and the different definitions are not directly related. We have also extended the simulations into the supersonic regime, up to convective Mach numbers of 10, where compressibility corrections are needed to reduce the growth rate of the mixing layer in accord with experiment (Figure \ref{f:chiravalle_comp}). Three common ``compressibility corrections'' from the literature (S89, Z90, and W92) perform very similarly and provide agreement with experimental results up to $M_c \approx 5$. The stress tensor modification implemented by GS11 provides similar results up to $M_c \approx 1$, but beyond this the model grows too slowly.

Three of the models tested (C06, GS11, and MS13) include buoyant effects (RT and RM instabilities). For these models, we use a simple stratified medium subject to constant acceleration to test the growth of the RT boundary layer. The model of GS11 shows the best agreement with experimental growth rates (Figure \ref{f:dimonte_growth}), while C06 grows too slowly and MS13 diverges at late times.

We then use the RANS models to simulate a generic astrophysical shock-cloud interaction. We follow the interaction in three dimensions for up to 10 cloud crushing times by implementing a co-moving grid. By using a consistent initial condition, we are able to compare global quantities as well as estimates of the mixing and injection returned by different turbulence models. We also generate an appropriate comparison by ensemble-averaging results from high-resolution inviscid simulations with grid-scale turbulence. We find that:
\begin{enumerate}
\item The $k$-$\varepsilon$ models of LS74 and MS13 and the $k$-$\omega$ model of W06 generate the least turbulence and corresponding lowest numerical viscosity.These models show the best agreement with the reference (TILES) result (Figure \ref{f:mdldiffplot}) at the fiducial resolution ($N_R = 25$).
\item The $k$-$L$ models of C06 and GS11 generate excessive turbulence within the cloud, leading to expansion, rapid disruption, and elevated mixing compared to the TILES result (Figure \ref{f:mdlcompare}). The W88 $k$-$\omega$ model generates excessive turbulence within the shock front, which also leads to enhanced disruption. Overall, the W88 and C06 models show the least agreement with the reference results (Figure \ref{f:mdldiffplot}).
\item Compressibility effects play a small role in the shock-cloud interaction, at least at the Mach number considered here ($M = 10$), as the compressibility corrections do not noticeably alter the simulation evolution or mixing estimates.
\item In agreement with previous work by P09, we show that the turbulence models are highly sensitive to the initial conditions (Figure \ref{f:icplot}). For large initial values of $k$ or $L$, the RANS models smooth the resolved dynamics beyond utility (Figure \ref{f:ictest_colr}); for small initial values, the RANS models have negligible effects.
\item Globally-averaged quantities vary only slightly with increasing resolution at resolutions higher than 25 cells per radius (Figure \ref{f:resplot}). While this agrees with previous work up to 100 cells per radius (PP16), we find that beyond this point turbulent mixing begins to be resolved [see also \ref{i:mix}] and thus alters the dynamics, preventing true convergence (Figure \ref{f:relerrplot}).
\item Estimates of the mixing decrease with increasing resolution up to 50 cells per radius (Figure \ref{f:resplot}), but beyond this point the mixing increases, up to a resolution of 200 cells per radius -- the current limit of our computational resources. This suggests that mixing in inviscid simulations does not trend toward zero at infinite resolution (Figure \ref{f:relerrplot}) but rather that the turbulent diffusivity becomes dominant when the numerical viscosity is sufficiently low. \label{i:mix}
\item The degree of mixing in the highest-resolution inviscid simulation ($N_R = 200$) agrees best with the predictions of the LS74 turbulence model (Figure \ref{f:mdlmixcompare}), but it is unknown what will occur at higher resolution or in a different application. Furthermore, the choice of numerical method (particularly the Riemann solver) can shift the mixing fraction in ILES simulations by nearly a factor of two (Figure \ref{f:sgtmresplot}).
\end{enumerate}

While the RANS turbulence models perform adequately in simple, specific test cases, it remains difficult to assess their veracity in complex dynamical applications. Further work toward understanding mixing in ILES simulations is necessary if proper calibrations are to be achieved.

\section*{Acknowledgements}
We thank the referee, Julian Pittard, for a constructive and insightful report. MDG thanks Jim Stone for a helpful discussion concerning turbulent mixing. Computations were performed on the KillDevil and Kure Clusters at UNC-Chapel Hill. We gratefully acknowledge support by NC Space Grant and NSF Grant AST-1109085.



\bibliographystyle{mnras}



\appendix

\section{Optimization and Performance}

The shock-cloud simulations were performed on the KillDevil Cluster at UNC Research Computing. To our knowledge, the run with $N_R = 200$ is the largest fixed-grid simulation of the three-dimensional shock-cloud interaction performed to date, with $4096\times2048\times2048$ grid cells. Evolving the simulation to $t=10 t_{\rm cc}$ required over 500,000~CPU-hours, with a maximum memory usage of nearly 13~TB across 2,048 CPUs. We built \textsc{Athena} using the Intel 13.1-2 compiler with the ``-O3'' optimization flag and the \textsc{MVAPICH2} 1.7 MPI library. Inter-process communications occurred over the QDR InfiniBand network.

Due to the fixed-grid nature of \textsc{Athena}, there is very little overhead in our simulations, and communication between processors is largely limited to transmission of boundary values after each update. \textsc{Athena} has been demonstrated to scale well out to 20,000 processors \citep{2008ApJS..178..137S}. We judge performance using the number of cells updated per CPU second. In our shock-cloud simulations, we find that the performance of the code is better for larger jobs, increasing from $2.02\times10^4$ cells per second at $N_R = 6$ up to $2.10\times10^5$ at $N_R = 200$. This increase is not surprising, as the ratio of computational work to inter-process communication increases with increasing resolution. In our largest simulation, the processors spent over 99\% of their time in active computation, indicating that the load is well-balanced and that inter-process communication over the InfiniBand network did not saturate significantly.


\bsp	
\label{lastpage}
\end{document}